# Soliton Shape and Mobility Control in Optical Lattices

**Yaroslav V. Kartashov[1], Victor A. Vysloukh[2], and Lluis Torner[1]**

*[1]ICFO-Institut de Ciencies Fotoniques, and Universitat Politecnica de Catalunya, Mediterranean Technology Park, 08860 Castelldefels (Barcelona), Spain*

*[2]Departamento de Fisica y Matematicas, Universidad de las Americas – Puebla, Santa Catarina Martir, 72820, Puebla, Mexico*

*E-mail: Yaroslav.Kartashov@icfo.es*

## Abstract

We present a progress overview focused on the recent theoretical and experimental advances in the area of *soliton manipulation* in optical lattices. Optical lattices offer the possibility to engineer and to control the diffraction of light beams in media with periodically-modulated optical properties, to manage the corresponding reflection and transmission bands, and to form specially designed defects. Consequently, they afford the existence of a rich variety of new families of nonlinear stationary waves and solitons, lead to new rich dynamical phenomena, and offer novel conceptual opportunities for all-optical shaping, switching and routing of optical signals encoded in soliton formats. In this overview, we consider reconfigurable optically-induced lattices as well as waveguide arrays made in suitable nonlinear materials. We address both, one-dimensional and multi-dimensional geometries. We specially target the new possibilities made possible by optical lattices induced by a variety of existing non-diffracting light patterns, we address nonlinear lattices and soliton arrays, and we briefly explore the unique features exhibited by light propagation in defect modes and in random lattices, an area of current topical interest and of potential cross-disciplinary impact.

# §1. Introduction

Nonlinear wave excitations are ubiquitous in Nature. They play an important role in our description of a vast variety of natural phenomena and their applications in many branches of science and technology, from optics to fluid dynamics and oceanography, solid-state physics, Bose-Einstein condensates (BECs), cosmology, etc. Examples are wide-spread (see, e.g., Infeld, Rowlands [1990]; Whitham [1999]; Moloney, Newell [2004]; Hasegawa, Matsumoto [1989]; Akhmediev, Ankiewicz [1997]; Kivshar, Agrawal [2003]; Stegeman, Segev [1999]).

Until recently, research in this area was commonly divided into two distinct, separated categories: Excitations in systems which can be described by continuous mathematical models and in the systems that can be effectively described by discrete equations. The difference between them is not merely quantitative, but qualitative, and often drastic. The paradigmatic example is the case of systems described by the nonlinear Schrödinger equation (NLSE). This is one of the few canonical equations that are encountered in many branches of nonlinear science when phenomena are studied at the proper scale length. The continuous version of NLSE belongs to that very special class of equations that in one-dimensional geometries are referred as completely-integrable and thus have rigorous single and multiple soliton solutions, which describe the stable and robust propagation of wave packets with an infinite number of conserved quantities. In contrast, systems modeled by the discrete NLSE may not conserve some of the fundamental quantities, such us the paraxial wave linear momentum, and thus feature a correspondingly restricted mobility. From a practical point of view, such qualitative differences are not necessarily advantageous or disadvantageous. For example, in the case of propagating light beams, the restricted mobility inherent to the discrete NLSE suggests new strategies for the routing or switching of light. By and large, the point is that the key differences existing between continuous and strictly-discrete mathematical models manifest themselves in the generation of correspondingly different physical phenomena.

However, there is a whole world in between systems modeled by totally continuous and totally discrete evolution equations. This world has been made accessible to experimental exploration in its full scope only recently, with the advent of optically-induced lattices in optics (see Fleischer, Segev, Efremidis, Christodoulides [2003]) and in Bose-Einstein conden-sates (see Morsch and Oberthaler [2006]). Tuning the strength of the lattices causes the system behavior to vary between a predominantly continuous model to a predominantly discrete one. The corresponding power concept termed "*tunable discreteness*" has direct important applications, e.g., for all-optical routing and shaping of light in optics, or for the generation and manipulation of quantum correlated matter waves in Bose-Einstein condensates. In addition to such applications, from a broad perspective, the ability to tune the strength of the genuinely discrete features of the nonlinear systems affords important applications in other areas of nonlinear science where the concept can be implemented. Optical lattices provide a unique laboratory to undertake such exploration. Here we present a progress overview focused on *soliton control*. We focus our attention on the salient possibilities afforded by the concept of tunable discreteness for soliton manipulation at large, and we refer the readers to other reviews for in-depth experimental advances (Stegeman, Christodoulides, Silberberg, Segev, Lederer, Assanto [Physics Reports]). Thus, we start by addressing the stationary properties of the different types of lattice solitons, and focus our attention on the dynamical properties by highlighting the main advances achieved to date. We address optical solitons, but most of the results hold as well for BECs.

The paper is organized as follows. In Sections 2 and 3 we discuss briefly the available nonlinear materials and structures that are currently used in most experiments with optical lattice solitons. In Section 4, we describe the experimental techniques available for optical lattice induction and mathematical models used to describe nonlinear light excitations in lattices. In Section 5 we address the salient properties of scalar and vector solitons supported by one-dimensional lattices. We discuss the impact on the soliton shape and soliton mobility caused by tunable static and dynamical optical lattices. We also address the potential stabilization of complex lattice soliton structures. In Section 6, the properties of solitons supported by two-dimensional optical lattices are considered. Addition of the second transverse coordinate remarkably affects properties of steady-state nonlinear waves. Optical lattices may have a strong stabilizing action for two-dimensional solitons and allow the existence of new types of solitons such as vortex solitons, possibilities that we discuss explicitly. In Section 7 lattices imprinted with different types of nondiffracting beams (including Bessel, Mathieu and parabolic ones) are discussed. Such lattices offer new opportunities for soliton managing and switching and constitute one of the key points to be highlighted in this area. In the absence of refractive index modulation three-dimensional solitons in self-focusing Kerr-type nonlinear media experience rapid collapse, but under appropriate conditions optical lattices may stabilize not only ground-state solitons but even higher-order three-dimensional solitons, even if the dimensionality of the lattice is lower than that of soliton, as briefly reviewed in Section 8. The properties of nonlinear optical lattices (including soliton arrays), together with the techniques that can be used to cause their stabilization, are addressed in Section 9. In Section 10 we review briefly the specific properties introduced by defect modes and by random lattices. Finally, in Section 11 the salient conclusions are given.

## §2.   Nonlinear materials

The availability and advent of suitable materials and fabrication techniques for the generation of optical lattices has been a key ingredient for the advancement of the field. In this section we briefly introduce the properties only of few optical materials which are routinely used for the experimental excitation of lattice solitons.

*Fused silica* is known to be a weakly nonlinear material ($n_2 \simeq 2.7 \times 10^{-16}$ cm$^2$/W). It has found applications in experiments with lattice solitons due to its unique possibility to write two-dimensional optical waveguides by tightly focusing femtosecond laser pulses generated by amplified Ti:Sa lasers. Permanent refractive index changes up to $1.3 \times 10^{-3}$ have been reached, with a typical spacing between 20 mm long waveguides of the order of $\sim 20\ \mu$m (Pertsch, Peschel, Lederer, Burghoff, Will, Nolte, Tünnermann [2004]; Szameit, Blomer, Burghoff, Schreiber, Pertsch, Nolte, Tunnerman [2005]). Formation of two-dimensional solitons in such waveguiding arrays was recently observed by Szameit, Burghoff, Pertsch, Nolte, Tünnermann, and Lederer [2006]. Evolution of linear light beams in arbitrary waveguide arrays, including laser-written ones, was addressed by Szameit, Pertsch, Dreisow, Nolte, Tünnermann, Peschel, and Lederer [2007].

Nonresonant nonlinearities in *semiconductor materials* may be orders of magnitude stronger than in silica. For instance, in AlGaAs at the wavelength 1.53 $\mu$m the nonlinearity coefficient amounts to $n_2 \sim 10^{-13}$ cm$^2$/W, and self-action effects are observable already for light intensities $\sim 10$ GW/cm$^2$. Such light intensities can be readily achieved with tightly focused pulsed radiation. Owing to highly advanced semiconductor processing technologies it is possible to fabricate individual semiconductor waveguides as well as waveguide arrays with the effective core areas $\sim 20~\mu\text{m}^2$. Such arrays were used to demonstrate discrete spatial solitons at power levels $\sim 500$ W (Eisenberg, Silberberg, Morandotti, Boyd, Aitchison [1998]). Waveguiding arrays fabricated from *polymer* inorganic-organic materials also have found important applications in the field of diffraction management (Pertsch, Zentgraf, Peschel, Brauer, and Lederer [2002]), and in the experimental observation of the optical analog of Zener tunneling (Trompeter, Pertsch, Lederer, Michaelis, Streppel, Brauer [2006]).

*Metal vapors* may feature resonant enhanced nonlinearities. Under proper conditions, e.g., in the absence of saturation, the nonlinear coefficient of sodium vapors for large frequency detuning is given by $n_2 = 8\pi^2 N \mu^4 / [n_0^2 c \hbar^3 (\omega - \omega_R)^3]$, where $N$ is the atomic concentration, $\mu$ is the dipole moment, and $\omega_R$ is the resonance frequency (Grischkowsky [1970]). In rubidium vapor, resonant enhancement is achieved close to the $D_2$ line at $\lambda = 780$ nm. The accessible values of the nonlinear coefficient range from $n_2 = 10^{-10}$ cm$^2$/W to $10^{-9}$ cm$^2$/W for typical concentrations $N \sim 10^{13}$ cm$^{-3}$. Self-trapping of a CW laser beam in sodium vapor was observed as early as 1974 (Bjorkholm, Ashkin [1974]). Saturation of the nonlinearity is typical for rubidium and sodium vapors. Then, the nonlinear contribution to the refractive index can be approximated as $\delta n = n_2 I / (1 + I / I_S)$, where $I_S$ is the saturation intensity. The nonlinearity sign, its strength, and saturation degree may be varied by changing the vapor concentration and by changing the detuning from resonance. Solitons were observed in metal vapors at power levels $\sim 100$ mW (Tikhonenko, Kivshar, Steblina, and Zozulya [1998]).

Nematic *liquid crystals* have emerged as suitable materials for experimentation with lattice solitons, thanks to their strong reorientational nonlinearity that may exceed the nonlinearity of standard semiconductors by several orders of magnitude (Simoni [1997]). Such crystals might be used for fabrication of periodic voltage-controlled waveguide arrays (Fratalocchi, Assanto, Brzdakiewicz, Karpierz [2004]). In such arrays the top and bottom cell interfaces provide planar anchoring of the liquid crystal molecules along the direction of light propagation. A set of periodically spaced electrodes (typical spacing $\sim 6~\mu\text{m}$) allows a bias to be applied across $\sim 5~\mu\text{m}$ thick crystal cell, thereby modulating the refractive index distribution thought molecular reorientation. The typical power required for spatial soliton formation in such arrays is $\sim 35$ mW at biasing voltage 1.2 V (see, e.g., recent reviews on solitons in nematic liquid crystal arrays by Brzdakiewicz, Karpierz, Fratalocchi, Assanto, Nowinowski-Kruszelnick [2005]; Assanto, Fratalocchi, Peccianti [2007], and section 5.7 of the present review).

*Photovoltaic nonlinearity* of photorefractive LiNbO crystal (see Bian, Frejlich, and Ringhofer [1997]), sets a rich playground for experiments with spatial solitons. For instance, self-trapping of an optical vortex in such crystals was reported by Chen, Segev, Wilson, Muller, and Maker [1997]. Two-dimensional *bright* photovoltaic spatial soliton have been observed in Cu:KNSBN crystal featuring focusing nonlinearity (She, Lee, and Lee [1999]). Photorefractive materials also can be used for fabrication of waveguide arrays. Such arrays might be created by titanium in-diffusion in a copper-doped LiNbO crystal where the optical nonlinearity (defocusing and saturable) arises from the bulk photovoltaic effect. Such waveguides combine high nonlinearity with adjustable contrast of linear refractive index. Typical LiNbO waveguide array consists of $4~\mu$m-wide titanium doped stripes separated by the same distance. Each channel forms a single-mode waveguide at 514.5 nm wavelength for refractive index modulation depth $\sim 3 \times 10^{-3}$, while the coupling constant $\sim 1~\text{mm}^{-1}$. The experimental observation of spatial gap solitons in such arrays at the power levels $\sim \mu$W was reported very recently (Chen, Stepic, Rüter, Runde, Kip, Shandarov, Manela, Segev [2005]).

An excellent material for experimentation with different types of nonlinear lattice waves is a doped photorefractive SBN crystal biased externally with a DC electric field. Due to a strong anisotropy of the electro-optic effect, the ordinary-polarized lattice-forming beams propagating almost linearly along the crystal are capable to induce relatively strong refractive index modulation for the extraordinary polarized probe beam, as it was suggested by Efremidis, Sears, Christodoulides, Fleischer, Segev [2002]. Recently it was demonstrated that nonlinear lattice waves in photorefractive SBN crystals might be observed at extremely low power levels of hundred nanowatts (Träger, Fisher, Neshev, Sukhorukov, Denz, Krolikowski, Kivshar [2006]).

Summarizing, there are a variety of materials suitable for the exploration of nonlinear wave propagation in periodic media. However, optical induction in suitable materials constitutes a real landmark advance for the purpose that we explore in this review, as it affords controlling the shape, strength and properties of the lattice with an unprecedented flexibility (see Section 4 below for a detailed discussion).

## §3.    Diffraction control in optical lattices

The ability to engineer and control diffraction opens broad prospects for light beam manipulation. In this section we discuss briefly a few milestones achieved in this area in lattices and waveguide arrays. In a linear bulk medium the spatial Fourier modes (plane waves) form an adequate set of eigenfunctions for the analysis of diffraction and refraction phenomena. Thus, diffraction of a collimated wave beam might be interpreted as dephasing of spatial harmonics that leads to convex wave front. Inside optical lattice the standard Fourier analysis becomes inefficient because of the energy exchange between an incident and Bragg-scattered waves. The adequate functional basis for description of wave propagation in such periodic environment is provided by the so-called Floquet-Bloch set of basis functions.

Figure 1 illustrates the dependencies of propagation constants of Bloch waves from the different bands on transverse wavenumber and the corresponding profiles of Bloch waves. In regions of convex band curvature the central mode propagates faster than its neighbors and the beam acquires a convex wave front during propagation. There are regions of normal diffraction where the wave behavior is fully analogous to that in the homogeneous media. By contrast, a group of modes in concave regions of band curvature evolve anomalously, producing a concave wave front. Thus, modification of input angle of the beam entering the periodic structure results in different rates and even signs of diffraction. Therefore, the effect of periodic lattice on propagation of laser beam depends to the great extent on the overall size of the input beam relative to the lattice period and on the depth of refractive index modulation.

### 3.1    Bloch waves

The linear interference of optical Bloch waves in periodically stratified materials was studied both theoretically and experimentally almost two decades ago (see Russell [1986] and references therein). More recently, Eisenberg, Silberberg, Morandotti, Aitchison [2000] proposed a scheme to design structures with controllable diffraction properties based on arrays of evanescently coupled waveguides. The scheme proposed is suitable to exploit the negative curvature of the diffraction curve of light in waveguide arrays in order to control diffraction. Arrays with reduced, canceled, and even reversed diffraction were technologically fabricated. Excitation of linear and nonlinear modes belonging to high-order bands of the Floquet-Bloch spectrum of periodic array was demonstrated by Mandelik, Eisenberg, Silberberg, Morandotti, Aitchison [2003]. To excite a single Bloch mode in a waveguide array the light beam was coupled at a grazing angle to the array from a region of a planar waveguide. Changing the tilt angle of the input beam allows selective excitation of periodic waves from different bands. Prism-coupling method for excitation of Floquet-Bloch modes and direct measurement of band structure of one-dimensional waveguide arrays was introduced by Rüter, Wisniewski, and Kip [2006]. Solitary waves bifurcating from Bloch-band edges in two-dimensional periodic media have been studied theoretically by Shi and Yang [2007].

In optically-induced photonic lattices, the Bragg scattering of light was studied experimentally in a biased photorefractive SBN:60 crystal (Sukhorukov, Neshev, Krolikowski, and Kivshar [2004]). This experiment showed the splitting of an input slant laser beam into several Bloch modes due to Bragg scattering. A novel, powerful experimental technique was recently presented for linear and nonlinear Brillouin zone spectroscopy of optical lattices (Bartal, Cohen, Buljan, Fleischer, Manela, Segev [2005]). This method relies on excitation of different Bloch modes with random-phase input beams and far-field visualization of emerging spatial spectrum. This technique facilitates mapping the borders of extended Brillouin zones and the areas of normal and anomalous dispersion. The possibility to acquire the full bandgap spectrum of photonic lattice of arbitrary profile was discussed by Fratalocchi and Assanto [2006c]. Nonlinear adiabatic evolution and emission of coherent Bloch waves in optical lattices was analyzed by Fratalocchi and Assanto [2007a]. Modulational instability of Bloch waves in one-dimensional saturable waveguide arrays was studied both theoretically and experimentally (Stepic, Rüter, Kip, Maluckov, and Hadzievski [2006]; Rüter, Wisniewski, Stepic, Kip [2007]).

Various types of two-dimensional Bloch waves were generated in a square photonic lattice by employing the phase imprinting technique (Träger, Fisher, Neshev, Sukhorukov, Denz, Krolikowski, and Kivshar [2006]). The unique anisotropic properties of lattice dispersion resulting from the different curvatures of the dispersion surfaces of the first and second spectral bands were demonstrated experimentally.

Nonlinear interactions between extended waves in optical lattices may lead also to new phenomena. For example, two Bloch modes launched into nonlinear photonic lattice evolve into a comb or a supercontinuum of spatial frequencies, exhibiting a sensitive dependence on the difference between the quasi-momenta of the two initially excited modes (see, Manela, Bartal, Segev, Buljan [2006]). Finally, it is worth to mention also that spatial four-wave mixing with Bloch waves was considered by Bartal, Manela, Segev [2006].

## 3.2   One- and two-dimensional waveguide arrays

The fabrication of simple GaAs waveguide arrays was achieved and analyzed more than three decades ago (see Somekh, Garmire, Yariv, Garvin, and Hunsperger [1973]). Channel waveguides were formed by the proton bombardment through a gold mask that allowed to create channels with width 2.5 $\mu$m separated by 3.9 $\mu$m and featuring relatively deep refractive index modulation $\sim 0.006$. Optical coupling accompanied by complete light transfer from the incident channel into adjacent ones was observed.

Since diffraction properties of light beams in lattices strongly depend on the propagation angle, tandems of different short segments of slant waveguide arrays (zigzag structure) might be used in order to achieve a desired average diffraction. This approach for diminishing the power requirement for lattice soliton formation was suggested by Eisenberg, Silberberg, Morandotti, Aitchison [2000], and it is closely linked with the idea of dispersion management from fiber optics technology. It was also shown that at higher powers, normal diffraction may lead to self-focusing and formation of bright solitons, but slight modification of the input tilt causing change of diffraction sign results in defocusing. Self-focusing and self-defocusing were achieved in the same medium, structure, and wavelength (Morandotti, Eisenberg, Silberberg, Sorel, and Aitchison [2001]).

Detailed experimental studies of anomalous refraction and diffraction of cw radiation in 1D array imprinted in inorganic-organic polymer was reported by Pertsch, Zentgraf, Peschel, Brauer, and Lederer [2002]. Discrete Talbot effects in 1D waveguide array was observed by Iwanow, May-Arrioja, Christodoulides, Stegeman, Min, Sohler [2005]. It was shown that in discrete configurations recurrence process is only possible for specific set of periodicities of the input patterns. Experimental demonstration of discrete Talbot effect was carried out in the array consisting of 101 waveguides imprinted in 70 mm long Z-cut LiNbO$_3$ wafer with lithography and titanium in-diffusion techniques.

Linear diffraction as well as nonlinear dynamics in 2D waveguide arrays can have more complicated character and can be much richer than those in 1D arrays (Hudock, Efremidis, Christodoulides [2004]; Pertsch, Peschel, Lederer, Burghoff, Will, Nolte, Tünnermann [2004]). For instance, diffraction in these arrays can be effectively altered by changing the beam's transverse Bloch vector orientation in the first Brillouin zone. In general, diffraction in 2D arrays can be made highly anisotropic and therefore permits the existence of elliptic solitons when nonlinearity comes into play. Under appropriate conditions not only the strength but also the sign of diffraction can differ for different directions (e.g., beam can experience normal diffraction in one direction and anomalous diffraction in the other one).

## 3.3    AM and FM transversely modulated lattices and arrays

In this subsection we discuss the new physical effects afforded by rather smooth (at the beam width scale) transverse modulation of the optical lattice parameters. As it was shown in the context of Bose-Einstein condensates, matter-wave soliton motion can be effectively managed by means of smooth variations of parameters of the optical lattice (Brazhnyi, Konotop, and Kuzmiak [2004]). Thus, linear, parabolic, and spatially-localized modulations of lattice amplitude and frequency were considered. For instance soliton could be accelerated, decelerated, or undergo reflection depending on the modulation function profile. The energy, effective mass, and soliton width can also be effectively controlled in such lattices. In this case the effective particle approximation provides a qualitative explanation of the main features of soliton dynamics, if soliton width substantially exceeds lattice period and lattice modulation is smooth enough.

Grating-mediated waveguiding was proposed by Cohen, Freedman, Fleischer, Segev, Christodoulides [2004]. Such a waveguiding is driven by a shallow 1D lattice with bell- or trough-shaped amplitude that slowly varies in the direction normal to the lattice wave-vector (lattice with amplitude modulation). Optical lattices with linear amplitude or frequency modulation was considered specifically for soliton control (Kartashov, Vysloukh, and Torner [2005d]). It was revealed, that the soliton trajectory can be controlled by varying the lattice depth and amplitude or frequency modulation rate. Also, it was uncovered that the effective diffraction of a light beam launched into the central channel of a lattice with a quadratic frequency modulation can be turned in strength and sign (Kartashov, Torner, and Vysloukh [2005]). Finally, complete suppression of linear diffraction in the broad band of spatial frequencies is accessible in FM lattices.

Recently, a direct visual observation of Bloch oscillations and Zener tunneling was achieved in 2D lattices photoinduced by four interfering plane waves in biased photorefractive crystal (Trompeter, Krolikowski, Neshev, Desyatnikov, Sukhorukov, Kivshar, Pertsch, Peschel, Lederer [2006]). In this experiment, a coordinate-dependent background illumination was used to create a refractive index distribution in the form of a periodic modulation superimposed onto the linearly increasing background.

Previously, it had been shown theoretically that Bloch oscillations can also emerge in discrete waveguides arrays with propagation constant linearly varying across the array (Peschel, Pertsch, Lederer [1998]). The occurrence of Bloch oscillations was demonstrated experimentally in an array of 25 AlGaAs waveguides with transversally varied refractive index and spacing (Morandotti, Peschel, Aitchison, Eisenberg, Silberberg [1999b]), as well as in thermo-optic polymer waveguide array with applied temperature gradient (Pertsch, Dannberg, Elflein, Brauer, Lederer [1999]). Photonic Zener tunneling between the bands of the polymer waveguide array was observed and investigated experimentally by Trompeter, Pertsch, Lederer, Michaelis, Streppel, Brauer [2006]. Such 1D polymer waveguide array was fabricated by ultra-violet lithography from an inorganic-organic polymer, while creating temperature gradient in the sample cause linear refractive index increase due to thermo-optic effect. Zener tunneling in 1D liquid crystal arrays was demonstrated by Fratallocchi, Assanto, Brzdakiewich, Karpierz, [2006] and Fratalocchi, Assanto [2006a]. A model allowing description of one-dimensional and more general two-dimensional Zener tunneling in two-dimensional periodic photonic structures and calculation of the corresponding tunneling probabilities was derived by Shchesnovich, Cavalcanti, Hickmann, Kivshar [2006]. Fratalocchi and Assanto [2007b] investigated nonlinear energy propagation in an optical lattice with slowly varying transverse perturbations. Symmetry breaking in such a system can provide a wealth of new phenomena, including nonreciprocal oscillations between Bloch bands and macroscopic self-trapping effects.

A binary waveguide array composed of alternating thick and thin waveguides was introduced by Sukhorukov and Kivshar [2002]. In such structure the effective refractive index experiences an additional transverse modulation. As a result, the existence of discrete gap solitons that possess the properties of both conventional discrete and Bragg grating solitons becomes possible. Such solitons and the effect of interband momentum exchange on soliton steering were observed experimentally in binary arrays fabricated in AlGaAs (Morandotti, Mandelik, Silberberg, Aitchison, Sorel, Christodoulides, Sukhorukov, and Kivshar [2004]).

The possibility to control the magnitude of dispersion experienced by a BEC wave-packet at the edges of spectral bands by modifying shape of double-periodic optical super-lattice was explored by Louis, Ostrovskaya, Kivshar [2005]. It was shown that an extra periodicity opens up additional narrow stopgaps in the band-gap spectrum, while the effective dispersion at the edges of these mini-gaps can be varied within much greater range than that accessible with single-period lattice.

Nonlinear light propagation and linear diffraction in disordered fiber arrays (or arrays with random spacing between waveguides) was investigated experimentally by Pertsch, Peschel, Kobelke, Schuster, Bartlet, Nolte, Tünnermann, Lederer [2004]. It was shown that for high excitation power diffusive spreading of light beam in such arrays is arrested by the focusing nonlinearity and formation of 2D discrete soliton is possible. The linear transmission and transport of discrete solitons in quasiperiodic waveguide arrays was studied by Sukhorukov [2006] and it was shown that under proper conditions dramatic enhancement of soliton mobility in such arrays is possible.

## 3.4    Longitudinally modulated lattices and waveguide arrays

Longitudinal modulation of diffractive/nonlinear properties of the transmitting medium is a powerful tool to control the light beam parameters. This technique can be extended to the case of optical lattices and waveguide arrays. A model governing the propagation of an optical beam in a diffraction managed nonlinear waveguide array with a step-like diffraction map was introduced by Ablowitz and Musslimani [2001]. This model allows the existence of discrete solitons whose width and peak amplitude evolve periodically. Its generalization to the case of vectorial interactions of two polarization modes propagating in diffraction managed array was developed later by Ablowitz and Musslimani [2002].

Interactions of diffraction managed solitons were studied in detail. Pertsch, Peschel, Lederer [2003] have considered soliton properties in inhomogeneous 1D waveguide arrays in the framework of a discrete model. It was shown that longitudinal periodic modulations of the coupling strength may lead to soliton oscillations and decay. Different types of optically induced dynamical lattices were considered, including lattices with simplest longitudinal amplitude modulation (Kartashov, Torner, Vysloukh [2004]), where amplification of soliton center swinging is possible, or more complex three-wave lattices (Kartashov, Torner, Christodoulides [2005]) that are capable to drag stationary solitons along predetermined paths (see also Sec. 6.6). Two-dimensional lattices evolving in longitudinal direction can be produced with several nondiffracting Bessel beams and it was shown that 2D solitons may undergo spiraling motion with controllable rotation rate in such lattices (Kartashov, Vysloukh, Torner [2005c]).

Linear and nonlinear light propagation in 1D waveguide array with a periodically bent axis was studied by Longhi [2005]. In the linear regime the analog of Talbot self-imagining effect was predicted. The dynamical localization of light in linear periodically curved waveguide arrays was observed experimentally (Longhi, Marangoni, Lobino, Ramponi, Laporta, Cianci, Foglietti [2006]; Iyer, Aitchison, Wan, Dignam, and de Sterke [2007]). A suitable periodic waveguide bending is capable to suppress discrete modulational instability of nonlinear Bloch waves. The dynamics of light in the presence of longitudinal defects of arbitrary extent in nonlinear waveguide arrays and uniform nonlinear media was studied by Fratalocchi, Assanto [2006b] and [2006d]. Dynamical super-lattices also may be used for manipulation with solitons as was suggested by Porter, Kevrekidis, Carretero-Gonzales, and Frantzeskakis [2006] for the case of 1D BECs. Some of soliton modes supported by optical lattices with localized defects can be driven across the lattice by means of the transverse lattice shift provided that this shift is performed with small enough steps (Brazhnyi, Konotop, Perez-Garcia [2006]). Recently it was predicted that Rabi-like oscillations and stimulated mode transitions occur with linear and nonlinear wave states in properly modulated waveguides and lattices (Kartashov, Vysloukh, Torner [2007a]). In particular, the possibilities of cascade stimulated transitions in systems supporting up to three modes of the same parity were shown. Such phenomenon occurs also in the nonlinear regime and in multimode systems, where the mode oscillation might be viewed as analogous to interband transitions.

## §4. Optically-induced lattices

The method of optical lattice induction is especially attractive because it allows creation of reconfigurable refractive index landscapes that can be fine-tuned by lattice-creating waves and easily erased in contrast to permanent, technologically fabricated waveguiding structures. Periodic, optically induced lattices might operate on both weakly and strongly coupled regimes between neighboring lattice guides depending on the intensities of waves inducing the lattice. This affords tunability of soliton behavior in such lattices from predominantly continuous to predominantly discrete, which is the core idea that we address in this review. The idea of optical lattice induction was put forward by Efremidis, Sears, Christodoulides, Fleischer, Segev [2002]. For completeness, in this section we discuss typical parameters of optical lattices that can be made with this method.

One or two-dimensional photoinduced lattice have to remain invariable along the propagation distance (typically, up to several centimeters). Experiments with photorefractive SBN crystal take advantage of its strong electro-optic anisotropy. The lattice-writing beams are polarized in ordinary direction, while the probe was polarized extraordinary along the crystalline c-axis. By applying a static electric field across this axis, the probe would experience photorefractive screening nonlinearity, while the lattice beams propagate in linear regime (Fleischer, Carmon, Segev, Efremidis, Christodoulides [2003]), that allows to achieve invariable lattice profile in the longitudinal direction. A further benefit of this system is that the sign of the nonlinearity (focusing/defocusing) can be altered by changing the polarity of the externally applied voltage.

## 4.1 Optical lattices induced by interference patterns

In the pioneering experiments, the periodic refractive index profile was photoinduced by interfering two ordinary polarized plane waves in a biased photorefractive SBN crystal (Fleischer, Carmon, Segev, Efremidis, Christodoulides [2003]). An external static electric field applied to the crystal creates periodic changes of the refractive index through the electro-optic effect. In SBN crystals orthogonally polarized waves features dramatically different electro-optic coefficients ( $r_{33} \simeq 1340 \text{ pm/V}$ , $r_{13} \simeq 67 \text{ pm/V}$ ), so that ordinary-polarized interfering plane waves propagate almost linearly and create the stable one-dimensional periodic lattice, whereas the extraordinary polarized probe beam experiences strong nonlinear self-action, which can be described through the nonlinear change of the refractive index $\delta n \sim -n_e^3 r_{33} E_0 / (I_{\text{dark}} + I_0 \cos^2(\pi x / d) + I)$, where $I_{\text{dark}}$ characterizes dark irradiance level, $I_0$ is the peak intensity of the lattice with the spatial period $d$, $I$ is the intensity of the probe beam, and $E_0$ is the static electric field. The intersection angle between plane waves determines lattice periodicity ($d \sim 10 \ \mu$m is a representative value for experiments with optically induced lattices), while experimentally achievable refractive index modulation depth is such lattice $\sim 10^{-3}$.

The propagation direction of the lattice-creating waves and probe beam may not necessarily coincide. Laser beams launched into tilted lattices may experience anomalous refraction and experience transverse displacement as demonstrated by Rosberg, Neshev, Sukhorukov, Kivshar, Krolikowski [2005]. In such experiment periodic lattice was induced in SBN:60 photorefractive crystal by interfering two ordinary polarized beams from a frequency-doubled Nd:YVO$_4$ laser at 532 nm. Variation in bias voltage alters refractive index modulation depth and bandgap lattice structure, which, in turn, modifies the diffraction properties of Bloch waves and allows to observe both positive or negative refraction of probe beams that selectively excite first of second spectral bands. Optical lattice might be also made partially incoherent and created by amplitude modulation rather than by coherent interference of multiple plane waves (Chen, Martin, Eugenieva, Xu, Bezryadina [2004]). Such lattices feature enhanced stability and might propagate even in weakly nonlinear regime due to suppression of incoherent modulation instability.

Optical induction also allows to produce square (Fleischer, Segev, Efremidis, and Christodoulides [2003]), hexagonal (Efremidis, Sears, Christodoulides, Fleischer, Segev [2002]), or triangular (Rosberg, Neshev, Sukhorukov, Krolikowski, and Kivshar [2007]) two-dimensional photonic lattices by interfering four or three plane waves. The representative example of lattice generated by four waves is shown in Fig. 2. Second dimension brings fundamentally new features into light propagation dynamics in periodic environment and allows a wealth of new lattice geometries.

Spatial light modulators allows to generate a rich variety of optical potentials, including nondiffracting ones, that might be used, for example, for guiding or trapping of atoms, as was suggested by McGloin, Spalding, Melville, Sibbett, Dholakia [2003]. Such modulators can be used as reconfigurable, dynamically controllable holograms and under proper conditions replace prefabricated micro-optical devices. For instance, current spatial light modulators operating in a phase-modulation regime offer $1024 \times 1024$ pixels with high diffraction efficiencies, which are comparable with etched holograms produced lithographically. The technique can be extended for optical lattice induction in photorefractive materials. Recently a programmable phase modulator was used for engineering of specific Bloch states from the second band of 2D optical lattice (Fisher, Träger, Neshev, Sukhorukov, Krolikowski, Denz, and Kivshar [2006]).

## 4.2 Nondiffracting linear beams

Nondiffractive linear light beams offer very attractive opportunities for photoinduction of steady-state photonic lattices of diverse topologies. The properties of solitons supported by photonic lattices imprinted by complex nondiffracting Bessel and Mathieu beams will be discussed in section 7. Here we briefly describe the most representative features of nondiffracting beams in linear homogeneous and periodic media.

Propagation of linear beams in uniform media is described by three-dimensional Helmholtz equation that admits separation into the transverse and longitudinal parts only in Cartesian, circular cylindrical, elliptic cylindrical, and parabolic cylindrical coordinates. Each of these coordinate systems then gives rise to a certain type of nondiffracting beam associated with a fundamental solution of Helmholtz equation in this coordinates. Details of experimental observation of parabolic, Bessel, and Mathieu beams can be found in papers by Durnin, Miceli, and Eberly [1987]; Gutierrez-Vega, Iturbe-Castillo, Ramirez, Tepichin, Rodrigues-Dagnino, Chavez-Cerda, and New [2001]; Lopez-Mariscal, Bandres, Gutierrez-Vega, and Chavez-Cerda [2005].

A powerful and universal representation of amplitude of any nondiffracting beam is provided by the reduced Whittaker integral:

$$Q(\zeta, \eta) = \int_{-\pi}^{\pi} G(\varphi) \exp\left[-ik\left(\zeta \cos\varphi + \eta \sin\varphi\right)\right] d\varphi, \tag{4.1}$$

where $G(\varphi)$ is the angular spectrum defined on the ring of radius $k$ in the frequency space. In the simplest case of Cartesian coordinates superposition of $M$ plane waves $G(\varphi) = \sum_{m=1}^{M} \delta(\varphi - \varphi_m)$ produces kaleidoscopic patterns (rectangular, honeycomb, etc). The angular spectrum $G(\varphi) = \exp(im\varphi)$ produces $m$-th order Bessel beams that represent fundamental nondiffracting solutions in circular cylindrical coordinate system, while $G(\varphi) = \mathrm{ce_m}(\varphi; a) + i\,\mathrm{se_m}(\varphi; a)$, with $\mathrm{ce_m}(\varphi; a)$ and $\mathrm{se_m}(\varphi; a)$ being angular Mathieu functions, produces $m$-th order Mathieu beams in elliptic coordinate system. Parabolic nondiffracting optical fields were discussed by Bandres, Gutierrez-Vega, and Chavez-Cerda [2004]. Some representative examples of transverse intensity distributions in most known types of nondiffracting beams are shown in Fig. 3. Using such beams in the technique of optical induction opens broad prospects for creation of the refractive index landscapes with novel types of symmetry. Besides this, nondiffracting beams find applications in diverse areas of physics, such as optical manipulation of small particles (McGloin and Dholakia [2005]), frequency doubling, or atom trapping.

Nondiffracting linear beams may also form in periodic media. Such beams were considered in two-dimensional periodic lattices (Manela, Segev, and Christodoulides [2005]). The links between the symmetry properties, phase structure of such beams and the number of spectral band that gives rise to the beam have been established. Recently, the possibility of diffraction-free propagation of localized light beams in materials with both transverse and longitudinal modulation of refractive index was predicted (Staliunas and Herrero [2006]; Staliunas, Herrero, and de Valcarcel [2006]; Staliunas and Masoller [2006]). In such materials the dominating (second) order of diffraction may vanish and thus the overall diffraction broadening of the beam is determined by the fourth-order dephasing of spatial Fourier modes.

### 4.3 Mathematical models of wave propagation

Several mathematical models are commonly accepted for description of soliton evolution in optical lattices. In the case of optical solitons all of them are based on the parabolic propagation equation for the slowly varying amplitude of light fields coupled to the material equation describing the corrections to the refractive index. Thus, in the simplest case of cubic nonlinear media their combination results in the canonical nonlinear Schrödinger equation describing the evolution of optical wave packets in the presence of the lattice, which in the case of one transverse dimension reads:

$$i\frac{\partial q}{\partial \xi} = -\frac{1}{2}\frac{\partial^2 q}{\partial \eta^2} + \sigma\,|q|^2\,q - pR(\eta)q. \qquad (4.2)$$

Here $q = (L_{\text{dif}} / L_{\text{nl}})^{1/2} A I_0^{-1/2}$ is the dimensionless complex amplitude of the light field; $A$ is the slowly varying envelope; $I_0$ is the input intensity; the transverse $\eta$ and longitudinal $\xi$ coordinates are normalized to the beam width $r_0$ and the diffraction length $L_{\text{dif}} = n_0 \omega r_0^2 / c$, respectively; $\omega$ is the carrying frequency; $n_0$ is the unperturbed refractive index; $L_{\text{nl}} = 2c / \omega n_2 I_0$; $\sigma = -1$ for focusing nonlinearity and $\sigma = +1$ in the case of defocusing nonlinearity; $p = L_{\text{dif}} / L_{\text{ref}}$ is the guiding parameter; $L_{\text{ref}} = c / \delta n \omega$; $\delta n$ is the refractive index modulation depth, while the function $R(\eta)$ describes the transverse refractive index profile. In the particular case of optically induced or technologically fabricated harmonic refractive modulation one can set $R(\eta) = \cos(\Omega \eta)$, where $\Omega$ is the lattice frequency. The generalization of the evolution equation (4.2) to the case of two transverse dimensions requires replacing $\partial^2 / \partial \eta^2$ by the two-dimensional transverse Laplacian $\partial^2 / \partial \eta^2 + \partial^2 / \partial \zeta^2$.

Note the analogy between the equations describing propagation of optical wavepackets and matter waves. Thus, Eq. (4.2) describes also dynamics of one-dimensional Bose-Einstein condensate confined in an optical lattice generated by a standing light wave of wavelength $\lambda$. In this case $q$ stands for the dimensionless mean-field wave function, the variable $\xi$ stands for time in units of $\tau = 2m\lambda^2 / \pi h$, with $m$ being the mass of the atoms and $h$ the Planck's constant, $\eta$ is the coordinate along the axis of the quasi-one-dimensional condensate expressed in units of $\lambda \pi^{-1}$. The parameter $p$ is proportional to the lattice depth $E_0$ expressed in units of recoil energy $E_{\text{rec}} = h^2 / 2m\lambda^2$. In quasi-one-dimensional condensates one has $\sigma = 2\lambda a_{\text{s}} N_{\text{a}} / \pi \ell^2$, where $a_{\text{s}}$ is the s-wave scattering length, $N_{\text{a}}$ is the number of atoms, and $\ell$ is the harmonic oscillator length.

The evolution equations typically admit of several conserved quantities. Thus, Eq. (4.2) conserves the total energy flow $U$ and the Hamiltonian $H$:

$$
\begin{aligned}
U &= \int_{-\infty}^{\infty} |q|^2 \, d\eta, \\
H &= \frac{1}{2} \int_{-\infty}^{\infty} (|\partial q / \partial \eta|^2 - 2pR|q|^2 + \sigma|q|^4) d\eta.
\end{aligned}
\tag{4.3}
$$

The method of creation of periodic lattices (especially two-dimensional ones) in nonlinear crystals generally relies on an optical induction technique (see section 3). With this technique optical lattice can be imprinted in a photosensitive crystal, for example, by inter-fering two or more plane waves. The interference pattern of several plane waves is a propagation invariant pattern and should not be affected by the nonlinearity of the crystal. At the same time the soliton beam released into the lattice should experience strong nonlinearity. Such conditions can be achieved in materials with a strong anisotropy of nonlinear response, e.g., in suitable photorefractive crystals. In this case, the lattice-creating waves and the soliton beam are polarized in orthogonal directions, so that nonlinearity affects only the soliton beam. In this configuration the lattice itself is not affected by the soliton beam, while the soliton beam does feel the periodic potential.

Since nonlinearity saturation is inherent for the photorefractive materials the model equation

$$i\frac{\partial q}{\partial \xi} = -\frac{1}{2}\frac{\partial^2 q}{\partial \eta^2} - \frac{Eq}{1 + S|q|^2 + R(\eta)}[S|q|^2 + R(\eta)] \tag{4.4}$$

is also frequently used for description of solitons in photorefractive optical lattices. Here $q = AI_0^{-1/2}$, $I_0$ is the input intensity; the normalization for transverse and longitudinal coordinate is similar to that for Eq. (4.2); $S = I_0/(I_{\text{dark}} + I_{\text{bg}})$ is the saturation parameter; $I_{\text{dark}}$ and $I_{\text{bg}}$ are dark and background radiation intensities; $R = I_{\text{latt}}/(I_{\text{dark}} + I_{\text{bg}})$, where $I_{\text{latt}}$ represents the intensity distribution in lattice-creating wave; $E = (1/2)(\omega r_0 n_0^2/c)^2 r_{\text{eff}} E_0$ is the dimensionless static biasing field applied to the crystal; $r_{\text{eff}}$ is the effective electro-optic coefficient corresponding to polarization of soliton beam. Under the assumptions $S|q|^2, R(\eta, \xi) \ll 1$, $E \gg 1$ the Eq. (4.4) can be transformed into the cubic nonlinear Schrodinger equation (4.2). It is important to stress that the depth of optically induced lattices can be tuned by changing intensities of lattice-creating waves. The model (4.4) generalized to the case of two transverse dimensions by replacing transverse Laplacian with its two-dimensional counterpart $\partial^2/\partial\eta^2 + \partial^2/\partial\zeta^2$ is also frequently used in the literature, though it describes strongly anisotropic and nonlocal response of such crystals only approximately (see section 9 where fully anisotropic model of photorefractive response is discussed).

Trivial-phase stationary solutions of Eq. (4.2) or (4.4) can be obtained in the form $q(\eta, \xi) = w(\eta)\exp(ib\xi)$, where $w(\eta)$ is a real function describing transverse profile, and $b$ is a real propagation constant. Thus, mathematically, families of lattice solitons are defined by the propagation constant $b$, lattice depth $p$, and particular lattice shape given by the function $R(\eta)$. Various families of soliton solutions can be obtained from a known family by using scaling transformations, which in the case of Eq. (4.2) are $q(\eta, \xi, p) \rightarrow \chi q(\chi \eta, \chi^2 \xi, \chi^2 p)$, where $\chi$ is the arbitrary scaling factor. The equation for soliton profiles obtained upon substitution of light field in such form into evolution equation can be solved numerically with standard relaxation or spectral methods. Analogously, split-step fast-Fourier method may be used to solve the very evolution equations with a variety of input conditions.

Stability of solitons can be tested by searching for perturbed solutions in the form $q = (w + u + iv) \exp(ib\xi)$, where $u(\eta, \xi)$ and $v(\eta, \xi)$ are real and imaginary parts of perturbation that can grow with a rate $\delta$ upon propagation. Substitution of light field in such form into Eq. (4.2), linearization around stationary solution $w$ and derivation of real and imaginary parts yields the following linear eigenvalue problem

$$
\begin{aligned}
\delta u &= -\frac{1}{2}\frac{\partial^2 v}{\partial \eta^2} + bv + \sigma w^2 v - pRv, \\
-\delta v &= -\frac{1}{2}\frac{\partial^2 u}{\partial \eta^2} + bu + 3\sigma w^2 u - pRu,
\end{aligned}
\tag{4.5}
$$

that can be solved numerically with standard linear eigenvalue solvers. The presence of perturbations corresponding to $\operatorname{Re}\delta \neq 0$ indicates an instability of the soliton solutions, while $\operatorname{Re}\delta \equiv 0$ indicates linear stability. The results of linear stability analysis may additionally be tested by direct numerical propagation of the soliton solutions perturbed with broadband input noise. The generalization of these techniques to the case of two transverse dimensions is straightforward.

## §5.    One-dimensional lattice solitons

In this section we describe the properties of scalar and vector solitons supported by one-dimensional lattices and briefly discuss the possibilities for control and manipulation of the soliton shape and width, its internal structure, and transverse mobility afforded by static and dynamical optical lattices with tunable parameters, imprinted in the materials with local and nonlocal nonlinear responses. We discuss the role of main factors that can lead to stabilization of complex lattice soliton structures.

Notice that the state of art in the field of spatial optical solitons in uniform nonlinear materials have been summarized in several books (Akhmediev and Ankiewicz [1997]; Trillo and Torruellas [2001]; Kivshar and Agrawal [2003]) and reviews (Stegeman and Segev [1999]; Kivshar and Pelinovsky [2000]; Torner [1998]; Etrich, Lederer, Malomed, T. Peschel and U. Peschel [2000]; Buryak, Di Trapani, Skryabin, and Trillo [2002]; Kivshar and Luther-Davies [1998]; Malomed, Mihalache, Wise, and Torner [2005]; Conti and Assanto [2004]). Lattice solitons are continuous counterparts of discrete solitons existing in waveguide arrays (for a recent reviews on discrete solitons see Aubry [1997]; Flach and Willis [1998]; Kevrekidis, Rasmussen, and Bishop [2001]; Christodoulides, Lederer, and Silberberg [2003]; Aubry [2006]). The discrete NLSE that is used to describe the evolution of nonlinear excitations in such systems has a rather universal character and can be used to describe light propagation even in tensorial systems and in the presence of nonparaxial and vectorial effects (Fratalocchi, Assanto [2007c]). As mentioned above, mathematically, the equations describing propagation of laser radiation in periodic media are analogous to that describing evolution of Bose-Einstein condensates in optical lattices, hence, many soliton phenomena predicted in nonlinear optics were encountered in BECs and vice versa (Dalfovo, Giorgini, Pitaevskii, and Stringari [1999]; Pitaevskii and Stringari [2003]; Abdullaev, Gammal, Kamchatnov, and Tomio [2005]; Morsch and Oberthaler [2006]).

## 5.1 Fundamental solitons

Lattice solitons form due to the balance of diffraction, refraction in the periodic lattice, and nonlinear self-phase-modulation. The existence of such solitons is closely linked to the band-gap structure of one-dimensional periodic lattice spectrum depicted in Fig. 1. Since the spectrum is composed of bands where only Bloch waves can propagate, and gaps where propagation of Bloch waves is forbidden, lattice solitons emerge as defect modes in gaps of lattice spectrum. The band-gap structure depends crucially on the lattice depth $p$, but in 1D it always includes a single semi-infinite gap and an unlimited number of finite gaps. The internal soliton structure is determined by the position of the soliton propagation constant inside the corresponding gap. In focusing media, the simplest fundamental (odd) solitons are formed in a semi-infinite gap (see Fig. 4(a) for an illustrative example of a soliton profile in the lattice $R(\eta) = \cos(\Omega\eta)$ with $\Omega = 4$ and focusing Kerr nonlinearity); the position of the intensity maximum for such soliton coincides with one of local lattice maximums. The properties of odd and other types of lattice solitons have been analyzed in different physical settings, including photorefractive optical lattices and BECs (Efremidis, Sears, Christodoulides, Fleischer, and Segev [2002]; Louis, Ostrovskaya, Savage, and Kivshar [2003]; Efremidis and Christodoulides [2003]; Kartashov, Vysloukh, and Torner [2004a]).

The monotonic increase of energy flow with propagation constant indicates stability of odd lattice solitons in accordance with the Vakhitov-Kolokolov stability criterion (Vakhitov and Kolokolov [1973]). Solitons existing near the gap edges transform into Bloch waves with the same symmetry. Besides odd solitons, optical lattices support also even solitons centered between neighboring maxima of $R(\eta)$ (Fig. 4(b)). In Kerr nonlinear media even solitons are unstable, while strong nonlinearity saturation may result in stabilization of even solitons accompanied by destabilization of odd ones (Kartashov, Vysloukh, and Torner [2004a]). Experimentally odd solitons in photorefractive lattices were observed by Fleischer, Carmon, Segev, Efremidis, and Christodoulides [2003] and by Neshev, Ostrovskaya, Kivshar, and Krolikowski [2003]. Figure 5 illustrates the experimental generation of odd and even lattice solitons.

Increasing the optical lattice depth inhibits localization of light in the vicinity of local lattice maxima, a process accompanied by the development of strong modulation of soliton profile. Under such conditions, one can employ the so-called *tight-binding* approximation to reduce Eq. (4.2) to a discrete NLSE obtained by Christodoulides and Joseph [1988] for arrays of evanescently coupled waveguides, where discrete solitons were observed by Eisenberg, Silberberg, Morandotti, Boyd, and Aitchison [1998]. Discrete matter-wave solitons were introduced by Trombettoni, Smerzi [2001].

Recently *dark discrete solitons* in one-dimensional arrays with defocusing nonlinearity have been also investigated (Smirnov, Rüter, Stepic, Kip, and Shandarov [2006]). Localized nonlinear dark modes displaying a phase jump in the center located either on-channel or in-between channels were observed, and the ability of the induced refractive index structures to guide light of a low-power probe beam was demonstrated. The interaction of a probe beam with dark and bright blocker solitons was studied in counterpropagating geometry by Smirnov, Rüter, Stepic, Shandarov, and Kip [2006]. Hadzievski, Maluckov, and Stepic [2007] presented a numerical analysis of the dynamics of dark breathers in lattices with saturable nonlinearity. Comparison of discrete dark solitons properties in lattices with saturable and Kerr nonlinearities was performed by Fitrakis, Kevrekidis, Susanto, and Frantzeskakis [2007]. Dark matter-wave solitons in optical lattices were discussed by Louis, Ostrovskaya, and Kivshar [2004]. The properties of dark solitons in dynamical lattices with cubic-quintic nonlinearity have been addressed recently by Maluckov, Hadzievski, and Malomed [2007].

## 5.2 Beam shaping and mobility control

Optical lattices offer rich opportunities for soliton control, by varying the lattice depth and period. In a lattice, at a given energy flow, the field amplitude necessary for soliton-like propagation amounts to lower values than that in homogeneous media. The width of lattice soliton also changes with increasing the lattice depth. More importantly, the transverse refractive index modulation profoundly affects the soliton mobility.

The simple effective particle approach (Scharf and Bishop [1993]; Kartashov, Zelenina, Torner, and Vysloukh [2004]) based on the equation

$$\frac{d^2}{d\xi^2}\langle \eta \rangle = \frac{p}{U}\int_{-\infty}^{\infty}|q|^2\frac{dR}{d\eta}\,d\eta \qquad (5.1)$$

for the integral soliton center $\langle \eta \rangle = U^{-1}\int_{-\infty}^{\infty}|q|^2\eta\,d\eta$, might be used to identify regimes of soliton propagation in the lattice. This approach requires substitution of a trial function for the light field, that can be chosen in the form $q = q_0\,\text{sech}[\chi(\eta - \langle \eta \rangle)]\exp[i\alpha(\eta - \langle \eta \rangle)]$, with $q_0$, $\chi$, and $\alpha$ being amplitude, inverse width, and incident angle, respectively. At small incident angles Eq. (5.1) predicts periodic oscillations of the soliton center inside the input channel. When the input angle exceeds a critical value $\alpha_{\text{cr}} = 2[p(\Omega\pi/2\chi)\sinh^{-1}(\Omega\pi/2\chi)]^{1/2}$ (associated with the height of the Peierls-Nabarro barrier created by the periodic potential (Kivshar and Campbell [1993])) the soliton leaves the input channel and starts traveling across the lattice, a process accompanied by radiative losses (Yulin, Skryabin, and Russell [2003]). The radiation may cause soliton trapping in one of the lattice channels. The controllable trapping in different lattice locations may find practical applications and have been studied in a variety of periodic systems, including discrete waveguide arrays (Aceves, De Angelis, Trillo, and Wabnitz [1994]; Krolikowski, Trutschel, Cronin-Golomb, Schmidt-Hattenberger [1994]; Aceves, De Angelis, T. Peschel, Muschall, Lederer, Trillo, Wabnitz [1996]; Krolikowski and Kivshar [1996]; Bang and Miller [1996]; Morandotti, Peschel, Aitchison, Eisenberg, Silberberg [1999]; Pertsch, Zentgraf, Peschel, Brauer, Lederer [2002]; Vicencio, Molina, Kivshar [2003]).

The transverse mobility of lattice solitons can be substantially enhanced in saturable media in the regime of strong saturation (Hadzievski, Maluckov, Stepic, Kip [2004]; Stepic, Kip, Hadzievski, Maluckov [2004]; Melvin, Champneys, Kevrekidis, Cuevas [2006]; Oxtoby and Barashenkov [2007]), an effect related to the stability exchange between even and odd solitons. Due to radiation emission by the moving solitons, soliton collisions in such systems are strongly inelastic (Aceves, De Angelis, Peschel, Muschall, Lederer, Trillo, Wabnitz [1996]; Meier, Stegeman, Silberberg, Morandotti, Aitchison [2004]; Meier, Stegeman, Christodoulides, Silberberg, Morandotti, Yang, Salamo, Sorel, Aitchison [2005]; Cuevas, Eilbeck [2006]).

## 5.3    Multipole solitons and soliton trains

Optical lattices can support scalar multipole solitons composed of several out-of-phase bright spots, when repulsive "forces" acting between spots in a bulk medium are compensated by the immobilizing action of the lattice (Fig. 4(c)). In discrete waveguide arrays such solitons are known as twisted strongly localized modes (Darmanyan, Kobyakov, Lederer [1998]; Kevrekidis, Bishop, Rasmussen [2001]). The important feature of multipole solitons in continuous lattices is that they become completely stable when their energy flow exceeds a certain threshold (Efremidis, Sears, Christodoulides, Fleischer, and Segev [2002]; Louis, Ostrovskaya, Savage, and Kivshar [2003]; Efremidis and Christodoulides [2003]; Kartashov, Vysloukh, and Torner [2004a]). Such stabilization takes place for multipole solitons of arbitrary higher order (soliton trains), containing multiple spots, provided that phase alternates by $\pi$ (in the case of focusing medium) between each two neighboring spots. This immediately suggests the important possibility to construct and to manipulate multi-peaked "soliton packets" beyond single "soliton bits", a feature that might open a new door in all optical-switching schemes. Notice that individual solitons can then be extracted or added into such trains.

The simplest dipole solitons in optical lattices were observed by Neshev, Ostrovskaya, Kivshar, and Krolikowski [2003]. He and Wang [2006] addressed specific dipole solitons residing completely in a single channel of a low-frequency optical lattice. Individual solitons and soliton trains have been also studied in lattices with competing cubic-quintic nonlinearities, where bistability was encountered for several soliton families (Merhasin, Gisin, Driben, Malomed [2005]; Wang, Ye, Dong, Cai, Li [2005]), as well as in pure quintic case (Alfimov, Konotop, and Pacciani [2007]). Periodic lattices support also truly infinite periodic waves that were studied in the context of BEC (for an overview, see Deconinck, Frigyik, and Kutz [2002]).

## 5.4    Gap solitons

The finite gaps of the Floquet-Bloch spectrum of the periodic lattice may also give rise to solitons, that are usually termed gap solitons. In contrast to solitons emerging from the semi-infinite gap (or total internal reflection gap), solitons from finite gaps are possible due to Bragg scattering from the periodic structure and exist in conditions of strong coupling between modes having opposite transverse wavevector components. Gap solitons exhibit multiple amplitude oscillations (Fig. 4(d)), while their internal symmetry depends on the particular gap that supports the solitons.

Gap-type excitations in discrete waveguide arrays were analyzed by Kivshar [1993]. One-dimensional gap solitons in continuous optically induced lattices were observed in photorefractive crystals by Fleischer, Carmon, Segev, Efremidis, and Christodoulides [2003]. Controlled generation and steering of gap solitons in optically induced lattices was reported by Neshev, Sukhorukov, Hanna, Krolikowski, Kivshar [2004]. Matter-wave gap solitons in Bose-Einstein condensates with repulsive inter-atomic interactions were observed by Eirmann, Anker, Albiez, Taglieber, Treutlein, Marzlin, Oberthaler [2004]. Gap solitons may also form in arrays of evanescently coupled waveguides fabricated in AlGaAs (Mandelik, Morandotti, Aitchison, Silberberg [2004]) and LiNbO (Chen, Stepic, Rüter, Runde, Kip, Shandarov, Manela, Segev [2005]; Matuszewski, Rosberg, Neshev, Sukhorukov, Mitchell, Trippenbach, Austin, Krolikowski, Kivshar [2005]).

The steering properties of gap solitons strongly depend on the sign and magnitude of the spatial group-velocity dispersion near the corresponding gap edge and may be anoma-lous, i.e. solitons may move in the direction opposite to the input tilt. Similarly to solitons forming in semi-infinite gap, solitons from finite gaps may form trains. In defocusing media, trains of in-phase gap solitons might be stable (Fig. 4(e)), while twisted modes (Fig. 4(f)) undergo strong exponential instabilities. Trains of gap solitons have been predicted in one-, two-, and three-dimensional optical lattices (Kartashov, Vysloukh, and Torner [2004a], Alexander, Ostrovskaya, Kivshar [2006]). Experimental observation of higher-order ode-dimensional gap solitons in defocusing waveguide arrays was presented by Smirnov, Rüter, Kip, Kartashov, and Torner [2007].

Gap soliton collisions are determined mostly by soliton localization. While broad low-energy solitons emerging close to the upper gap edge may interact almost elastically, the collisions of high-power gap solitons is inelastic and may lead to soliton fusion (Dabrowska, Ostrovskaya, Kivshar [2004]; Malomed, Mayteevarunyoo, Ostrovskaya, Kivshar [2005]). The interaction between two well-localized parallel solitons in one-dimensional discrete saturable systems has been investigated using defocusing lithium niobate nonlinear waveguide arrays by Stepic, Smirnov, Rüter, Prönneke, Kip, and Shandarov [2006]. Gap solitons have been studied in more complicated but still periodic systems such as binary waveguide arrays (Sukhorukov, Kivshar [2002]) or quasiperiodic optical lattices (Sakaguchi, Malomed [2006]).

Gap solitons are typically stable in the deep of the gap, and near the band edge they feature specific oscillatory instabilities similar to those observed in fiber Bragg gratings (Barashenkov, Pelinovsky, Zemlyanaya [1998]). Weak instabilities of gap solitons arise due to resonant energy redistributions between different gaps, while perturbation eigenmodes associated with such instability are poorly localized (Pelinovsky, Sukhorukov, Kivshar [2004]).

Motzek, Sukhorukov, and Kivshar [2006] studied the dynamical reshaping of polychromatic beams in periodic and semi-infinite photonic lattices and the formation of polychromatic gap solitons. The propagation of polychromatic light in nonlinear photonic lattices has been reviewed recenetly by Sukhorukov, Neshev, and Kivshar [2007]. In particular, the observation of localization of supercontinuum radiation in waveguide arrays imprinted in $LiNbO_3$ crystals and the possibilities for dynamical control of the output spectrum were reported by Neshev, Sukhorukov, Dreischuh, Fischer, Ha, Bolger, Bui, Krolikowski, Eggleton, Mitchell, Austin, and Kivshar [2007].

## 5.5  Vector solitons

Vectorial interactions (cross-phase modulation) between several light fields may significantly enrich the dynamics of soliton propagation in the periodic structure. The propagation of two mutually incoherent beams in the lattice with focusing ($\sigma = -1$) or defocusing ($\sigma = +1$) Kerr nonlinearity can be described by the system of equations:

$$i\frac{\partial q_{1,2}}{\partial \xi} = -\frac{1}{2}\frac{\partial^2 q_{1,2}}{\partial \eta^2} + \sigma q_{1,2}(|q_1|^2 + |q_2|^2) - pR(\eta)q_{1,2} \qquad (5.2)$$

The model (5.2) can be enlarged to take into account coherent interaction between orthogonally polarized beams in birefringent media, when four-wave-mixing is taken into account. Vectorial coupling between strongly localized modes in discrete waveguide arrays generates new soliton families having no counterparts in the scalar case (Darmanyan, Kobyakov, Schmidt, Lederer [1998]; Ablowitz, Musslimani [2002]). The simplest vector solitons were observed in AlGaAs arrays in the presence of four-wave-mixing (Meier, Hudock, Christodoulides, Stegeman, Silberberg, Morandotti, Aitchison [2003]).

In lattices, vectorial interaction between solitons emerging from the same gap stabilizes beams that are unstable when propagating alone. Thus, even solitons in focusing media are stabilized when coupled with twisted mode, and vice-versa, in defocusing media in-phase trains stabilize their twisted counterparts (Kartashov, Zelenina, Vysloukh, Torner [2004]). Localization of two-component Bose-Einstein condensates and formation of bright-brigth and dark-bright matter-wave solitons in optical lattices was considered by Ostrovskaya, Kivshar [2004a].

Coupling between solitons emerging from the neighboring gaps of lattice spectrum may also lead to formation of composite vector states as predicted by Cohen, Schwartz, Fleischer, Segev, Christodoulides [2003]; Sukhorukov, Kivshar [2003]. Importantly, one of the components in such vector states is always unstable alone due to its symmetry dictated by the gap number. Transient interband mutual focusing and defocusing of waves exhibiting diffraction of different magnitude and sign was observed by Rosberg, Hanna, Neshev, Sukhorukov, Krolikowski, Kivshar [2003] in optical lattices. The coherent interactions between two beams closely resembling profiles of Bloch waves from two neighboring bands also results in formation of nonlinear localized excitations or breathers that experience beating upon propagation in the array (Mandelik, Eisenberg, Silberberg, Morandotti, Aitchison [2003]). The impact of nonlinearity saturation on properties of vector solitons in discrete waveguide arrays was studied by Fitrakis, Kevrekidis, Malomed, and Frantzeskakis [2006], while experimental observation of vector solitons in saturable discrete waveguide arrays was performed by Vicencio, Smirnov, Rüter, Kip, and Stepic [2007].

## 5.6    Soliton steering in dynamical lattices

All lattices considered so far were invariant along the direction of propagation. Variation of the lattice shape in the longitudinal direction opens a wealth of new opportunities for soliton control. Several types of dynamical lattices were considered in the literature, including lattices with monotonic or periodic variation of depth and width of guiding channels as well as lattices with periodically curved channels. In particular, a longitudinal variation of coupling strength in modulated waveguide array can resonantly excite internal modes of discrete solitons, which may lead to soliton breathing, splitting, or transverse motion depending on the symmetry of the excited mode (Peschel, Lederer [2002]; Pertsch, Peschel, Lederer [2003]).

Self-imaging at periodic planes may occur in arrays of periodically curved waveguides, while modulational instability is inhibited under conditions of self-imaging (Longhi [2005]). Periodic waveguide arrays built of several tilted segments support breathing discrete diffraction managed scalar and vector solitons reproducing their shapes on each period of the structure (Ablowitz, Musslimani [2001]; Ablowitz, Musslimani [2002]). A longitudinal modulation causes parametric amplification of soliton swinging inside lattice channel that may be used to detect submicron soliton displacements (Kartashov, Torner, Vysloukh [2004]). Modulated lattices might be used for stopping and trapping moving solitons, while initially stationary solitons can be transferred to any prescribed position by a moving lattice (Kevrekidis, Frantzeskakis, Carretero-Gonzalez, Malomed, Herring, Bishop [2005]; Dabrowska, Ostrovskaya, Kivshar [2006]). Gap solitons undergo abrupt delocalization and compression in optical lattice with varying strength (Baizakov, Salerno [2004]).

An interesting opportunity to control output soliton position was encountered in dynamical optical lattices that exhibit a finite momentum in the transverse plane. Such lattices can be induced by three plane waves $A \exp(\pm i\alpha\eta)\exp(-i\alpha^2\xi/2)$, $B\exp(i\beta\eta)\exp(-i\beta^2\xi/2)$, where $a,b$ are wave amplitudes and $\pm\alpha,\beta$ are propagation angles, so that lattice profile is given by $R(\eta,\xi)=4A^2\cos^2(\alpha\eta)+B^2+4AB\cos(\alpha\eta)\cos[\beta\eta+(\alpha^2-\beta^2)\xi/2]$. In this case the transverse momentum can be transferred from the lattice to the soliton beam thus leading to a controllable drift (Kartashov, Torner, Christodoulides [2005]; Garanovich, Sukhorukov, Kivshar [2005]). Perturbation theory based on the inverse scattering transform might be used to calculate the tilt $\phi$ (or instantaneous propagation angle) of the soliton beam $q(\eta,\xi)=\chi\,\mathrm{sech}[\chi(\eta-\phi\xi)]\exp[i\phi\eta-i(\chi^2-\phi^2)\xi/2]$ in three-wave lattice. One gets

$$\phi=\frac{2\pi AB}{\chi}\left(\frac{\beta-\alpha}{(\beta+\alpha)\sinh[\pi(\beta-\alpha)/2\chi]}-\frac{\beta+\alpha}{(\beta-\alpha)\sinh[\pi(\beta+\alpha)/2\chi]}\right)\left(1-\cos\frac{\beta^2-\alpha^2}{2}\xi\right),$$

(5.3)

Thus, the drift grows linearly with the amplitude $B$ of the third plane wave. All-optical beam steering in modulated photonic lattices induced by three-wave interference was observed experimentally by Rosberg, Garanovich, Sukhorukov, Neshev, Krolikowski, Kivshar [2006] and is illustrated in Fig. 6. Soliton steering and fission in optical lattices that fade away exponentially along the propagation direction was considered by Kartashov, Vysloukh, and Torner [2006]. Garanovich, Sukhorukov, and Kivshar [2007] discussed phenomenon of nonlinear light diffusion in periodically curved arrays of optical waveguides. Soliton steering by a single continuous wave that dynamically induces a photonic lattice, as well as interactions of several solitons in the presence of such waves were addressed by Kominis, Hizanidis [2004] and Tsopelas, Kominis, Hizanidis [2006].

### 5.7 Lattice solitons in nonlocal nonlinear media

Under appropriate conditions the nonlinear response of materials can be highly nonlocal, a phenomenon that drastically affects the propagation of light. New effects attributed to nonlocality were encountered, e.g., in photorefractive media, liquid crystals and thermo-optical media (see Krolikowski, Bang, Nikolov, Neshev, Wyller, Rasmussen, Edmundson [2004] for a review). Nonlocality of the nonlinear response also strongly affects properties of lattice solitons. Propagation of a laser beam in nonlocal media with an imprinted transverse refractive index modulation can be described by the equation:

$$i\frac{\partial q}{\partial \xi} = -\frac{1}{2}\frac{\partial^2 q}{\partial \eta^2} - q\int_{-\infty}^{\infty} G(\eta - \lambda)|q(\lambda)|^2 \, d\lambda - pR(\eta)q, \qquad (5.4)$$

where $G(\eta)$ is the response function of the nonlocal medium. In the limit $G(\eta) \to \delta(\eta)$ one recovers the case of local response, while strongly nonlocal medium is described by slowly decaying kernels $G(\eta)$. Symmetric strongly nonlocal response reduces symmetry-breaking instabilities of lattice solitons, including even solitons in focusing and twisted gap modes in defocusing media. The Peierls-Nabarro potential barrier for soliton moving across the lattice is reduced due to nonlocality, an effect that results in the corresponding enhancement of the transverse soliton mobility (Xu, Kartashov, Torner [2005a]). By tuning the lattice depth one can control the mobility of lattice solitons in materials with asymmetric nonlocal response (Kartashov, Vysloukh, Torner [2004b], Xu, Kartashov, Torner [2006]).

Liquid crystals belong to the rare class of nonlinear materials simultaneously featuring intrinsically nonlocal nonlinearity and allowing for formation of tunable refractive index landscapes (Peccianti, Conti, Assanto, De Luca, Umeton [2004]; Fratalocchi, Assanto, Brzdakiewicz, Karpierz [2004], Fratalocchi, Assanto [2005]). Figure 7 shows setup and experimentally observed lattice solitons in periodically biased nematic liquid crystal. The lattice depth is tuned by varying the voltage applied to periodically spaced electrodes on the liquid crystal cell. A similar technique was used for demonstration of multiband vector breathers (Fratalocchi, Assanto, Brzdakiewicz, Karpierz [2005a]). The experimental observation of power-dependent beam steering over angles of several degrees in a voltage-controlled array of channel waveguides defined in a nematic liquid crystal via electrooptic reorientational response was observed by Fratalocchi, Assanto, Brzdakiewicz, Karpierz [2005b]. Such steering was demonstrated for light beams at mW power levels. An overview of the latest experimental and theoretical advances in investigation of discrete light propagation and self-trapping in nematic liquid crystals is given in Fratalocchi, Brzdakiewicz, Karpierz, Assanto [2005].

## §6.   Two-dimensional lattice solitons

In this section we describe the properties of solitons supported by two-dimensional optical lattices. By and large, adding the second transverse coordinate profoundly affects properties of self-sustained nonlinear excitations. In particular, two dimensional solitons in uniform Kerr nonlinear media may experience catastrophic collapse in contrast to their one-dimensional counterparts (Berge [1998]). Optical lattices typically play strong stabilizing action for two-dimensional solitons and allow for existence of new types of solitons with unusual internal structures. Here we describe the conditions required for existence of stable two-dimensional lattice solitons. Two-dimensional settings also give rise to vortex solitons that are impossible in lower-dimensional settings. The properties of optical and matter-wave vortices are described in details in a recent review by Desyatnikov, Torner, Kivshar [2005]. In this section we address the existence and specific features of vortex solitons in periodic media. Various types of two-dimensional solitons have been studied in discrete nonlinear systems (for a recent reviews see Aubry [1997]; Flach and Willis [1998]; Hennig, Tsironis [1999]; Kevrekidis, Rasmussen, and Bishop [2001]; Christodoulides, Lederer, and Silberberg [2003]; Aubry [2006]; as well as the focus issues Physica D **119**, 1 (1998) and Physica D **216**, 1 (2006) devoted to discrete solitons in nonlinear lattices). Recent developments in the field of two-dimensional matter-wave lattice solitons are summarized by Kevrekidis, Carretero-Gonzalez, Frantzeskakis, Kevrekidis [2004]; Morsch and Oberthaler [2006].

## 6.1 Fundamental solitons

Similarly to the case of one-dimensional solitons addressed in section 5, the properties of two-dimensional solitons in optical lattices are dictated by the band-gap structure of the lattice spectrum. However, in clear contrast with the 1D case, bands in the spectrum of 2D lattice can strongly overlap, thereby considerably restricting the number of gaps in the spectrum or even completely eliminating all finite gaps. This is the main property that distinguishes 2D lattices from their 1D counterparts possessing an infinite number of finite gaps. The number of finite gaps in the spectrum of 2D lattice is dictated by the particular lattice shape, depth, and period.

The simplest 2D lattice solitons in focusing media emerge from the semi-infinite gap (Efremidis, Sears, Christodoulides, Fleischer, Segev [2002]; Yang, Musslimani [2003]; Efremidis, Hudock, Christodoulides, Fleischer, Cohen, Segev [2003]; Baizakov, Malomed, Salerno [2003]). The absolute intensity maximum for such solitons coincides with one of the local lattice maxima. In Kerr media, an increase of the soliton amplitude at a fixed lattice depth is accompanied by a gradual energy flow concentration within single lattice channel, while the soliton profile approaches that of unstable Townes soliton. When $b$ approaches the lower edge of the semi-infinite gap, the soliton amplitude decreases, a process that is accompanied by soliton expansion over many lattice periods (transition into corresponding Bloch wave).

In Kerr nonlinear media periodic refractive index modulation results in collapse suppression and stabilization of fundamental soliton almost in the entire domain of its existence, except for the narrow region near the lower edge of semi-infinite gap, where energy flow of 2D solitons diverges in contrast to energy flow of their 1D counterparts (Musslimani, Yang [2004]). Fundamental 2D lattice solitons were observed in a landmark experiment conducted in optically induced lattice in photorefractive SBN crystal (Fleischer, Segev, Efremidis, Christodoulides [2003]). Photorefractive crystals possess strong electro-optic anisotropy, so that optical lattices produced by four ordinarily polarized interfering plane waves propagate in the linear regime, while extraordinarily polarized probe beam experience a significant screening nonlinearity (see section 3). Since the nonlinear correction to refractive index is determined by the total intensity distribution, the lattice creates the periodic environment invariable in propagation direction where the probe beam may experience discrete diffraction or form a lattice soliton when nonlinearity is strong enough (see Fig. 8).

Photonic lattices might be induced with partially incoherent light (Martin, Eugenieva, Chen, Christodoulides [2004]; Chen, Martin, Eugenieva, Xu, Bezryadina [2004]). In this case the lattice is created by amplitude modulation rather than by coherent interference of multiple waves and might be operated in either linear or nonlinear regimes (the latter regime is achieved when lattice is extraordinarily polarized) due to suppression of incoherent modulational instabilities and allow for observation of 2D fundamental solitons and soliton trains. Quasi-1D (i.e., uniform in one transverse direction and periodic in other direction) optical lattices created in a bulk medium with a pair of interfering plane waves also support quasi-1D or 2D solitons.

In focusing media, quasi-1D lattice solitons are transversally unstable, developing a snake-like shape at low powers or experiencing neck-like break-up at high powers, as observed by Neshev, Sukhorukov, Kivshar, Krolikowski [2004]. However, the periodic refrac-tive index modulation inhibits the development of the transverse instability. Note that the modulational instability of discrete solitons in coupled waveguides with group velocity dispersion may lead to formation of mixed spatio-temporal quasi-solitons (Yulin, Skryabin, and Vladimirov [2006]). Fully localized 2D solitons in quasi-1D lattices can be completely stable in the most part of their existence domain in both Kerr and saturable media. Such solitons can be set into radiationless motion in the direction where lattice is uniform, which opens a wealth of opportunities for studying tangential soliton collisions (Baizakov, Malomed, Salerno [2004]; Mayteevarunyoo, Malomed [2006]). Weakly transversally modulated quasi-1D lattices support localized 2D solitons even in defocusing media (Ablowitz, Julien, Musslimani, Weinstein [2005]). Two-dimensional solitons have been also thoroughly studied in various discrete systems (see for example Pouget, Remoissenet, Tamga [1993]; Mezentsev, Musher, Ryzhenkova, Turitsyn [1994]; Kevrekidis, Rasmussen, Bishop [2000] and reviews mentioned above).

Mobility of two-dimensional solitons in both continuous and discrete system can be strongly enhanced in the regime of nonlinearity saturation (Vicencio, Johansson [2006]). Specific optical discrete X-waves are possible in normally dispersive nonlinear waveguide arrays with focusing nonlinearity. Such waves can be excited for a wide range of input conditions while their properties strongly depart from the properties of X-waves in bulk or waveguide configurations (Droulias, Hizanidis, Meier, Christodoulides [2005]). Discrete X-waves were recently observed in AlGaAs arrays by Lahini, Frumker, Silberg, Droulias, Hizanidis, Morandotti, Christodoulides [2007].

## 6.2    Multipole solitons

Multipole solitons in 2D optical lattices may have much richer shapes than their 1D counterparts. The basic properties of such solitons were investigated by Musslimani, Yang [2004]; Kartashov, Egorov, Torner, Christodoulides [2004]. Multipole solitons in focusing media are characterized by a $\pi$ phase jump between neighboring spots (2D lattices also support even solitons comprising two in-phase spots, but such solitons are typically unstable). Multipole solitons exist above some minimal energy flow determined by the lattice parameters. In contrast to fundamental or to even solitons, multipole solitons do not transform into delocalized Bloch waves at the cutoff for existence and their existence domain does not occupy the whole semi-infinite gap.

Lattices stabilize multipole solitons even in Kerr nonlinear media when the energy flow (or $b$) exceeds a certain critical value. In principle, the number of solitons that can be incorporated into the multipoles is not limited, but the energy threshold for stabilization grows monotonically with the number of poles. This suggests the possibility of packing a number of individual solitons with properly engineered phases into one stable matrix to encode complex soliton images and letters. Experimental observation of dipole solitons was conducted by Yang, Makasyuk, Bezryadina, Chen [2004a] and [2004b] in optical lattice induced with partially coherent light beams in photorefractive crystals.

Higher-order stationary necklace solitons were experimentally generated by launching vortex beams into a partially coherent lattice (Yang, Makasyuk, Kevrekidis, Martin, Malomed, Frantzeskakis, Chen [2005]). This effect results in the generation of octagonal necklace soliton featuring $\pi$ phase difference between adjacent spots (see Fig. 9). Notice that for the observation of formation of necklace-like structures in optically induced lattices, Yang, Makasyuk, Kevrekidis, Martin, Malomed, Frantzeskakis, Chen [2005] fixed the intensity of the input beam and increased the voltage applied to the crystal, something that results in a simultaneous increase of the lattice and nonlinearity strengths, in contrast to conventional method of observation of transition between linear discrete diffraction and nonlinear self-trapping upon increase of the intensity of the input beam at fixed applied voltage (see, e.g., Fleischer, Segev, Efremidis, Christodoulides [2003]).

It should be pointed out that even more complex soliton trains can be generated when relatively broad stripe beam is launched into 2D lattice (Chen, Martin, Eugenieva, Xu, Bezryadina [2004]). In this case, due to highly anisotropic character of the nonlinear response of the crystal, increasing the nonlinearity facilitates soliton train formation or, vise versa, enhances discrete diffraction for orthogonal orientations of the input stripe.

An experimental study of the dynamics of off-site excitations conducted by Lou, Xu, Tang, Chen, Kevrekidis [2006] has shown that single beam launched between two sites of periodic photonic lattice excites asymmetric lattice soliton when nonlinearity reaches a threshold. This is an indication that the branch corresponding to stable asymmetric lattice solitons may bifurcate from the unstable even soliton branch at high powers. Multipole solitons in two-dimensional discrete system have been also thoroughly studied (see, for exam-ple, paper by Kevrekidis, Malomed, Bishop [2001] and references therein). Extended structures in nonlinear discrete systems occupying multiple channels were addressed by Kevrekidis, Gagnon, Frantzeskakis, and Malomed [2007].

## 6.3    Gap solitons

If the parameters of a 2D lattice are chosen in such a way that the lattice spectrum possesses at least single finite gap, conditions are set to form 2D gap solitons. Such solitons were observed in photorefractive optically induced lattices by Fleischer, Segev, Efremidis, Christodoulides [2003]. A probe beam was launched at an angle with respect to the lattice plane to ensure that the beam experiences anomalous diffraction, while the polarity of biasing field was chosen to produce defocusing nonlinearity. As a result, self-trapped wave packet was observed with a staggered phase structure, where phase changes by $\pi$ between neighboring lattice channels. A theoretical analysis of the properties of 2D gap solitons shows that they always exhibit power thresholds for their existence, and they might be stable provided that the propagation constant is not too close to the gap edges (Ostrovskaya, Kivshar [2003]; Efremidis, Hudock, Christodoulides, Fleischer, Cohen, Segev [2003]). Such solitons can form bound states, whose stability properties are defined by the nonlinearity sign. For example, in defocusing media the in-phase combinations of 2D gap solitons are stable, while out-of-phase combinations (or twisted solitons) may be prone to symmetry-breaking instabilities. Stability of gap solitons was addressed numerically (Richter, Motzek, and Kaiser [2007]). Experimental observation of gap soliton trains in a two-dimensional defocusing photonic lattices was presented by Lou, Wang, Xu, Chen, and Yang [2007]. Peleg, Bartal, Freedman, Manela, Segev, and Christodoulides [2007] reported on specific conical diffraction and gap soliton formation in honeycomb photorefractive lattices. Dipole-like two-dimensional gap solitons in defocusing optically induced lattices were observed by Tang, Lou, Wang, Song, Chen, Xu, Chen, Susanto, Law, and Kevrekidis [2007].

Arrays of 2D gap solitons might be generated via Bloch wave modulational instabilities (Baizakov, Konotop, Salerno [2002]; Brazhnyi, Konotop, Kuzmiak [2006]). Novel types of localized beams supported by the combined effects of total internal and Bragg reflection in two-dimensional lattices were recently observed by Fischer, Träger, Neshev, Sukhorukov, Krolikowski, Denz, Kivshar [2006]. Such gap states originate from the X-symmetry point of the lattice spectrum and possess a reduced symmetry and highly anisotropic diffraction and mobility properties. Hybrid semi-gap spatio-temporal solitons may exist in quasi-1D lattices, where localization in space is achieved due to interplay of lattice, diffraction, and defocusing nonlinearity, while localization in time is due to defocusing nonlinearity and normal group velocity dispersion (Baizakov, Malomed, Salerno [2006a]). Higher-order solitons in two-dimensional discrete lattices with defocusing nonlinearity were recently analyzed by Kevre-kidis, Susanto, and Chen [2006]. In-band (or embedded) solitons that appear in trains and that bifurcate from Bloch modes at the interior high-symmetry X points within the first band of lattice spectrum were observed in 2D photonic lattices with defocusing nonlinearity [Wang, Chen, Wang, Yang [2007]).

## 6.4    Vector solitons

In contrast to their 1D counterparts, 2D vector solitons in periodic lattices have been studied only for the case of incoherent interactions between soliton components. Experimental observation of the simplest 2D vector soliton in optically induced partially coherent photonic lattice was performed by Chen, Bezryadina, Makasyuk, Yang [2004]; Chen, Martin, Eugenieva, Xu, Yang [2005]. It was demonstrated that two mutually incoherent beams can lock into a fundamental vector soliton while propagating along the same lattice site, although each beam alone would experience discrete diffraction under similar conditions. The components $w_{1,2}$ of such soliton feature similar functional shapes and can be considered as $\phi$-projections of the profile of scalar lattice soliton $w(\eta, \zeta)$, i.e., $w_1 = w \cos \phi$ and $w_1 = w \sin \phi$.

When two mutually incoherent beams are launched into neighboring lattice sites, they form dipole-like vector soliton featuring strong mutual coupling between components in both poles. Lattices support also vector solitons formed by two dipole field components in the case of orthogonal orientation of dipoles in different components. The total intensity distribution in such solitons is reminiscent to that for quadrupole scalar solitons, but stability of vector states is enhanced in comparison with their scalar counterparts (Rodas-Verde, Michinel, Kivshar [2006]). Lattices can also support 2D vector solitons whose components emerge from the same or different finite gaps of the lattice spectrum. Such solitons were studied by Gubeskys, Malomed, Merhasin [2006] for the case of repulsive interspecies and zero intraspecies interactions in the context of a binary BECs. It was found that intragap solitons that are typically bound states of tightly and loosely bound components originating from first and second gaps can be found in deep enough lattices and are stable in a wide regions of their existence.

Vectorial interactions in photonic lattices optically induced in photorefractive crystals result in the formation of counterpropagating mutually incoherent self-trapped beams analyzed by Belic, Jovic, Prvanovic, Arsenovic, Petrovic [2006]. Optical lattices may play a strong stabilizing role for such counterpropagating solitons (Koke, Träger, Jander, Chen, Neshev, Krolikowski, Kivshar, Denz [2007]). The time-depended rotation of couterpropagating mutually incoherent self-trapped beams in optically induced lattices was discussed by Jovic, Prvanovic, Jovanovic, and Petrovic [2007]. Interactions of counterpropagating discrete solitons in a photorefractive waveguide array were studied experimentally by Smirnov, Stepic, Rüter, Shandarov, and Kip [2007]. Finally, the existence and stability of strongly localized 2D vectorial modes in discrete arrays were analyzed by Hudock, Kevrekidis, Malomed, Christodoulides [2003].

## 6.5   Vortex solitons

Vortex solitons are characterized by a special beam shape, in amplitude and in phase, that carries a nonzero orbital angular momentum, which is related to energy circulation inside the vortex. Thus, vortex solitons realize higher-order, excited states of the corresponding nonlinear systems. Vortex solitons carry screw phase dislocations located at the points where the intensity vanishes. In uniform focusing media vortex solitons feature localized ring-like intensity distribution. However, they are highly prone to azimuthal modulational instabilities that result in their spontaneous self-destruction into ground-state solitons (for a review on vortex solitons see Desyatnikov, Torner, Kivshar [2005]).

The properties of vortex solitons in periodic media differ dramatically from the properties of their radially symmetric counterparts in uniform materials. The profiles of lattice vortex solitons have the form $q(\eta, \zeta, \xi) = [w_{\mathrm{r}}(\eta, \zeta) + i w_{\mathrm{i}}(\eta, \zeta)] \exp(i b \xi)$, where $w_{\mathrm{r}}$ and $w_{\mathrm{i}}$ are real and imaginary parts of the complex field $q$. The topological winding number $m$ (or dislocation strength) of such complex scalar field can be defined by the circulation of the field phase $\arctan(w_{\mathrm{i}} / w_{\mathrm{r}})$ around the phase singularity. The result is an integer. Substitution of the vortex field in such a form into the nonlinear Schrodinger equation in the simplest case of Kerr nonlinearity yields

$$-\frac{1}{2}\left(\frac{\partial^2}{\partial \eta^2} + \frac{\partial^2}{\partial \zeta^2}\right)w_\mathrm{r} + bw_\mathrm{r} + \sigma(w_\mathrm{r}^2 + w_\mathrm{i}^2)w_\mathrm{r} - pR(\eta, \zeta)w_\mathrm{r} = 0,$$
$$-\frac{1}{2}\left(\frac{\partial^2}{\partial \eta^2} + \frac{\partial^2}{\partial \zeta^2}\right)w_\mathrm{i} + bw_\mathrm{i} + \sigma(w_\mathrm{r}^2 + w_\mathrm{i}^2)w_\mathrm{i} - pR(\eta, \zeta)w_\mathrm{i} = 0,$$

(6.1)

where $\sigma = -1$ corresponds to focusing nonlinearity and $\sigma = 1$ corresponds to defocusing nonlinearity, while $R(\eta, \zeta)$ stands for the profile of the lattice. Linear stability analysis for lattice vortex solitons can be conducted by considering perturbed solutions in the form $q = (w_\mathrm{r} + U + iw_\mathrm{i} + iV)\exp(ib\xi)$, where $U = u(\eta, \zeta)\exp(\delta\xi)$, $V = v(\eta, \zeta)\exp(\delta\xi)$ are real and imaginary parts of perturbation that can grow with a complex rate $\delta$ upon propagation. The linearized equations for perturbation components then become

$$\delta u = -\frac{1}{2}\left(\frac{\partial^2}{\partial \eta^2} + \frac{\partial^2}{\partial \zeta^2}\right)v + bv + 2\sigma w_\mathrm{i}w_\mathrm{r}u + \sigma(3w_\mathrm{i}^2 + w_\mathrm{r}^2)v - pRv,$$
$$-\delta v = -\frac{1}{2}\left(\frac{\partial^2}{\partial \eta^2} + \frac{\partial^2}{\partial \zeta^2}\right)u + bu + \sigma(3w_\mathrm{r}^2 + w_\mathrm{i}^2)u + 2\sigma w_\mathrm{i}w_\mathrm{r}v - pRu.$$

(6.2)

Absence of perturbations satisfying Eq. (6.2) for $\mathrm{Re}\,\delta > 0$ indicates vortex stability.

Vortex lattice solitons were initially introduced in discrete systems (Aubry [1997]; Johansson, Aubry, Gaididei, Christiansen, Rasmussen [1998]; Malomed, Kevrekidis [2001]; Kevrekidis, Malomed, Bishop, Frantzeskakis [2001]; Kevrekidis, Malomed, Chen, Frantzeskakis [2004]). Detailed theoretical analysis of properties and stability of vortex solitons in continuous lattices induced in focusing Kerr media was performed by Yang, Musslimani [2003]. It was shown that the even simplest lattice vortices exhibit strongly modulated intensity distributions with four pronounced intensity maxima whose positions almost coincide with positions of lattice maxima (Baizakov, Malomed, Salerno [2003]; Yang, Musslimani [2003]). In focusing medium such vortices originate from the semi-infinite gap of the lattice spectrum. Phase changes by $\pi/2$ between neighboring vortex spots. Two types of charge-1 vortices were found: Off-site and on-site. Contrary to naive expectations, the intensity distribution for high-amplitude vortices exhibits four very well localized bright spots concentrated in the vicinity of lattice maxima. Both, on-site and off-site vortex solitons that are oscillatory unstable near lower cutoff become completely stable above certain critical value $b_{cr}$. Note that in the isotropic photorefractive model widely used in the literature, vortex solitons feature a limited stability domain since strong nonlinearity saturation destabilizes them (Yang [2004]).

Experimental observation of off-site vortices in photorefractive optical lattices was performed by Neshev, Alexander, Ostrovskaya, Kivshar, Martin, Makasyuk, Chen [2004], while Fleischer, Bartal, Cohen, Manela, Segev, Hudock, Christodoulides [2004] have reported the generation of both on-site and off-site vortex configurations. Typically, vortex lattice solitons can be excited with input ring-like beams carrying screw phase dislocation of unit charge. The interaction of vortex soliton with surrounding photonic lattice can modify the vortex structure. In particular, nontrivial topological transformations such as flipping of vortex charge and inversion of its orbital angular momentum are possible (Bezryadina, Neshev, Desyatnikov, Young, Chen, and Kivshar [2006]).

Periodic media impose important restrictions on the available charges of vortex solitons. Using general group theory arguments Ferrando, Zacares, Garcia-March [2005]; Ferrando, Zacares, Garcia-March, Monsoriu, Fernandez de Cordoba [2005], demonstrated that unlike in homogeneous medium, no symmetric vortices of arbitrary high order (or charge) can be generated in 2D nonlinear systems possessing a discrete-point rotational symmetry. In particular, in the case of square optical lattices produced by interference of four plane waves, the realization of discrete symmetry forbids the existence of symmetric vortex solitons with charges higher than two. Still, dynamical rotating quasi-vortex double-charged solitons, reversing the topological charges and the direction of rotation through a quadrupole-like transition state can exist in two-dimensional photonic lattices (Bezryadina, Eugenieva, Chen [2006]). Stable stationary vortex solitons with topological charge 2 introduced by Oster and Johansson [2006] incorporate eight main excited lattice sites and feature rhomboid-like intensity distribution. Vortex solitons in discrete-symmetry systems are shown to behave as angular Bloch modes characterized by an angular Bloch momentum (Ferrando [2005]). Ideally, symmetric periodic lattices can support asymmetric vortex solitons, including rhomboid, rectangular, and triangular vortices (Alexander, Sukhorukov, Kivshar [2004]). Such nonlinear localized structures describing elementary circular flows can be analyzed using energy-balance relations. Higher-order asymmetric vortex solitons and multipoles containing more than four lobes on a square lattice were constructed by Sakaguchi, Malomed [2005].

Finite gaps of the lattice spectrum may give rise to specific gap vortex solitons carrying vortex-like phase dislocation and existing due to the Bragg scattering. Vortices originating from the first gap were studied in defocusing media by Ostrovskaya, Kivshar [2004b]. Such vortices experience delocalizing transition akin to flat-phase gap solitons when their energy flow becomes too high. Due to their gap nature, such vortex solitons develop very unusual chessboard-like phase structure. The process of dynamical generation of spatially localized gap vortices was studied by Ostrovskaya, Alexander, Kivshar [2006]. The anisotropic nature of the photorefractive nonlinearity may substantially affect the properties of gap vortex solitons propagating in such media (see Richter and Kaiser [2007]). In focusing media vortex lattice solitons arising from the X symmetry points and residing in the first finite gap of lattice spectrum were suggested by Manela, Cohen, Bartal, Fleischer, Segev [2004]. Such solitons have the phase structure of a counterrotating vortex array and are stable for moderate lattice depths and soliton intensities. Figure 10 illustrates such gap vortex that was experimentally observed by Bartal, Manela, Cohen, Fleischer, Segev [2005]. Lattices with more complicated geometries support vortex solitons featuring unusual symmetries, such as multivortex solitons in triangular lattices (Alexander, Desyatnikov, and Kivshar [2007]). Kartashov, Ferrando, and García-March [2007] predicted the existence of a new type of discrete-symmetry lattice vortex solitons that can be considered as a coherent state of dipole solitons carrying a nonzero topological charge.

Vectorial interactions between lattice vortices substantially enriches their propagation dynamics. For example composite states of vortex solitons originating from different gaps can be stable in certain parameter regions, while the decay of unstable representatives of such family may be accompanied by exchange of angular momentum between constituents and transformation into other type of stable composite soliton (Manela, Cohen, Bartal, Fleischer, Segev [2004]). Some novel scenarios of charge-flipping instability of incoherently coupled on-site and off-site vortices were encountered by Rodas-Verde, Michinel, Kivshar [2006]. Counterpropagating vortices in optical lattices are also subject of current studies (Petrovic [2006]; Petrovic, Jovic, Belic, and Prvanovic [2007]).

## 6.6    Topological soliton dragging

Two-dimensional settings give rise to a new class of lattices possessing higher topological complexity, e.g., regular lattices containing dislocations. In contrast to lattices that feature regular shapes, the guiding properties of lattices with dislocations can drastically change in the vicinity of dislocation, a possibility that leads to new phenomena. Such lattices may be produced, e.g., by interference of plane wave and wave carrying one or several nested vortices (Kartashov, Vysloukh, Torner [2005b]). In the simplest case of single-vortex wave lattice profile is given by $R(\eta, \zeta) = |\exp(i\alpha\eta) + \exp(im\phi + i\phi_0)|^2$, where $\alpha$ is the plane wave propagation angle, $\phi$ is the azimuthal angle, $m$ is the vortex charge, and $\phi_0$ defines orientation of vortex origin. Such lattices are not stationary and distort upon propagation, but they can be implemented in two-dimensional condensates strongly confined in the direction of light propagation.

Representative profiles of lattices produced by interference of plane wave and wave with nested vortex are shown in Fig. 11. They feature clearly pronounced channels where solitons can travel. The screw phase dislocation results in appearance of the characteristic fork, so that some channels may fuse or disappear in the dislocation. Notice that lattice periodicity is broken only locally, in a close proximity of dislocation. The higher the vortex charge, the stronger the lattice topological complexity and more channels fuse or disappear in the dislocation. Lattice intensity gradient in the vicinity of dislocation results in appearance of forces acting on the soliton, that lead to soliton's fast transverse displacement when it is launched into one of lattice channels in the vicinity of dislocation (Figs. 11(a),(b)). The dependence of sign and magnitude of forces acting on solitons on the lattice topology makes it possible to use combinations of several dislocations (produced by spatially separated oppositely charged vortices) for creation of double lattice defects (Figs. 11(c),(d)), or soliton traps, which feature equilibrium regions that are capable of supporting stationary solitons. Positive traps form when several lattice rows fuse to form the region where $R(\eta, \zeta)$ is locally increased, while negative traps feature breakup of several lattice channels.

The effective particle approach (see section 5.2) yields the equations

$$\frac{d^2}{d\xi^2}\langle\eta\rangle = \frac{p}{U}\int\int_{-\infty}^{\infty}|q|^2\frac{dR}{d\eta}\,d\eta d\zeta, \quad \frac{d^2}{d\xi^2}\langle\zeta\rangle = \frac{p}{U}\int\int_{-\infty}^{\infty}|q|^2\frac{dR}{d\zeta}\,d\eta d\zeta, \quad (6.3)$$

for soliton center coordinates $\langle \eta \rangle, \langle \zeta \rangle$, which allow one to construct the map of forces $F_\eta \sim d^2 \langle \eta \rangle / d\xi^2$ and $F_\zeta \sim d^2 \langle \zeta \rangle / d\xi^2$ acting on solitons. Such maps built for simplest Gaussian ansatz for the soliton profile are shown in Figs. 11(e),(f) for both positive and negative traps. Both traps repel all outermost solitons (some of the possible soliton positions are indicated by circles in this figure) but support stationary solitons located inside traps, which are stable for positive traps and unstable for negative traps. This result confirms the crucial importance of lattice topology for properties and stability of solitons that it supports. Notice, that solitons in two-dimensional lattices possessing defects, dislocations, and quasicrystal structures were also investigated in details by Ablowitz, Ilan, Schonbrun, Piestun [2006].

## §7.    Solitons in Bessel lattice

The simplest optical lattices exhibit a square or honeycomb shape. Lattices with different types of symmetry also exist, and offer new opportunities for soliton existence, managing, and control. Such lattices can be imprinted with different types of nondiffracting beams, including Bessel beams, parabolic beams, or Mathieu beams. All these beams are directly associated with the fundamental nondiffracting solutions of the wave equation in circular, parabolic, or elliptical cylindrical coordinates and thus exhibit a different symmetry. The techniques to generate various types of nondiffracting beams can be found in Durnin, Miceli, Eberly [1987]; Gutierrez-Vega, Iturbe-Castillo, Ramirez, Tepichin, Rodrigues-Dagnino, Chavez-Cerda, New [2001]; Lopez-Mariscal, Bandres, Gutierrez-Vega, Chavez-Cerda [2005]. In this section we describe the properties of solitons supported by lattices imprinted by radially symmetric and azimuthally modulated Bessel beams, and by Mathieu beams.

### 7.1    Rotary and vortex solitons in radially symmetric Bessel lattices

Radially symmetric Bessel lattices can be imprinted optically with zero-order or higher order Bessel beams, whose field distribution can be written as $q(\eta, \zeta, \xi) = J_n[(2b_{\text{lin}})^{1/2} r] \exp(in\phi) \exp(-ib_{\text{lin}} \xi)$, where $r = (\eta^2 + \zeta^2)^{1/2}$, $\phi$ is the azimuthal angle, $n$ is the charge of the beam, and the parameter $b_{\text{lin}}$ sets the transverse extent of the beam core. Such beams preserve their transverse intensity distribution upon propagation in a linear medium. Rigorously speaking, such beams are infinitely extended and thus carry infinite energy. However, accurate approximations can be generated experimentally in a number of ways. Bessel lattices can be imprinted in photorefractive crystals with the technique used for the generation of harmonic patterns by incoherent vectorial interactions (Efremidis, Sears, Christodoulides, Fleischer, and Segev [2002]) and in Bose-Einstein condensates. The simplest Bessel lattices, that exhibit an intensity given by the zero-order Bessel beam, contain a central guiding core surrounded by multiple bright rings. Lattices produced by higher-order beams have zero intensity at $r = 0$.

The field of the radially symmetric soliton in such lattice can be written as $q(\eta, \zeta, \xi) = w(r) \exp(im\phi) \exp(ib\xi)$, where $m$ is the topological charge. This yields the following equation for soliton profile in the case of Kerr nonlinear response:

$$\frac{1}{2}\left(\frac{d^2w}{dr^2} + \frac{1}{r}\frac{dw}{dr} - \frac{m^2w}{r^2}\right) - bw - \sigma w^3 + pR(r)w = 0. \tag{7.1}$$

To analyze the linear dynamical stability of the stationary soliton families, one has to search for perturbed solutions in the form $q(\eta, \zeta, \xi) = [w(r) + u(r, \xi)\exp(in\phi) + v^*(r, \xi)\exp(-in\phi)]\exp(ib\xi + im\phi)$, where the perturbation components $u, v$ could grow with complex rate $\delta$ upon propagation, $n$ is the azimuthal perturbation index. Linearization of nonlinear Schrodinger equation around a stationary solution yields the eigenvalue problem:

$$\begin{aligned} i\delta u &= -\frac{1}{2}\left(\frac{d^2}{dr^2} + \frac{1}{r}\frac{d}{dr} - \frac{(m+n)^2}{r^2}\right)u + bu + \sigma w^2(v + 2u) - pRu, \\ -i\delta v &= -\frac{1}{2}\left(\frac{d^2}{dr^2} + \frac{1}{r}\frac{d}{dr} - \frac{(m-n)^2}{r^2}\right)v + bv + \sigma w^2(u + 2v) - pRv. \end{aligned} \tag{7.2}$$

In the case of focusing nonlinear media ($\sigma = -1$), the simplest solitons are located in the central core of the Bessel lattice (Kartashov, Vysloukh, Torner [2004c]). At low energy flows such solitons are wide and extend over several lattice rings, while at high energies they are narrow and concentrate mostly within the core of the lattice. Solitons in lattices produced by higher-order Bessel beams exhibit an intensity deep at $r = 0$. Importantly, Bessel lattice imprinted in Kerr nonlinear media can suppress collapse and thus stabilize radially symmetric solitons in the most part of their existence domain.

A fascinating example of localized structure supported by Bessel lattices is given by solitons trapped in different rings of the lattice. Such solitons suffer from exponential instabilities close to low-amplitude cutoff where they are broad, but they are completely stable above a threshold $b$ value. Solitons trapped inside the rings can be set into rotary motion by launching them tangentially to the lattice rings. The important point is that such solitons do not radiate upon rotation. A rich variety of interaction scenarios is encountered for collective motion of solitons inside the lattice rings (Fig. 12).

Interactions between in-phase solitons located in the same ring may cause their fusion into a high-amplitude slowly rotating soliton, while collision of out-of-phase solitons results in formation of steadily rotating soliton pairs with a fixed separation between components. Collisions of out-of-phase solitons in different rings may be accompanied by reversal of their rotation direction after each act of collision. Notice that radially symmetric Bessel lattices may also support multipole solitons composed of out-of-phase spots residing in different lattice rings.

Experimental demonstration of soliton formation in optically induced ring-shaped radially symmetric photonic lattices was performed by Wang, Chen, Kevrekidis [2006]. The transition from diffraction (symmetric in the case of excitation of central guiding core and asymmetric in the case of excitation of one of the outer lattice rings) to guidance of extraordinarily polarized probe beam by the lattice was observed by tuning the lattice and nonlinearity strengths. In addition to immobile solitons trapped in the central lattice core and in different lattice rings, Wang et al observed controlled rotation of solitons launched at different angles into lattice rings (see Fig. 13).

The polarity of the biasing electric field applied to the photorefractive crystal can be selected in such way that Bessel-like beams with a bright central core induce photonic lattices featuring a decreased refractive index in the center. Such ring lattice with a low-refractive-index core surrounded by concentric rings (akin to a photonic band-gap fiber) may guide linear beams whose localization is achieved due to reflection on outer rings of the lattice. Guidance of linear beam in Bessel-like low-refractive-index core lattice was demonstrated experimentally by Wang, Chen, Yang [2006].

Solitons supported by ring-shaped radially-periodic lattices were studied by Hoq, Kevrekidis, Frantzeskakis, Malomed [2005] and by Baizakov, Malomed, Salerno [2006b]. In contrast to Bessel lattices whose intensity slowly decreases at $r \to \infty$, radially-periodic lattices do not decay at infinity, a property that substantially affects soliton behavior and stability domains. In particular, no localized states exist in such lattices in the linear limit and all localized states are truly nonlinear. In focusing media, radially-periodic lattices support fundamental and higher-order (i.e., concentrated in outer rings) axially uniform solitons. In such settings all higher-order solitons are azimuthally unstable. On the contrary, solitons of the same type may be stable in defocusing media. Lattices with defocusing nonlinearity also give rise to radial gap solitons.

Surface waves localized at the edge of radially-periodic guiding structures consisting of several concentric rings are also possible (Kartashov, Vysloukh, Torner [2007b]). Such surface waves rotate steadily upon propagation and, in contrast to nonrotating waves, for high rotation frequencies they do not exhibit power thresholds for their existence. There exists an upper limit for the surface wave rotation frequency, which depends on the radius of the outer guiding ring and on its depth.

Bessel optical lattices imprinted in defocusing media with higher-order radially symmetric nondiffracting Bessel beams can also support stable ring-profile vortex solitons (Kartashov, Vysloukh, Torner [2005a]). Such solitons exist because Bessel lattice compensates defocusing and diffraction, thereby affording confinement of light that is impossible in the uniform defocusing medium. In the low-energy limit, vortex solitons transform into linear guided modes of the Bessel lattice, while high-energy vortices greatly expand across the lattice and acquire multi-ring structure. Since modulational instability is suppressed in defocusing media ring-shaped vortices in Bessel lattices may be stable. The higher the soliton topological charge the deeper the lattice modulation necessary for vortex stabilization.

Rotating quasi-one-dimensional and two-dimensional guiding structures are also interesting. In the limit of fast rotation light beams propagating in such structures feel the average refractive index distribution that may be equivalent, for example, to the refractive index distribution optically-induced by the Bessel beams. Solitons supported by two-dimensional rotating lattices with focusing nonlinearity were introduced by Sakaguchi and Malomed [2007]. Rotating lattices, even quasi-one-dimensional, were found to support localized ground-state and vortex solitons in defocusing media provided that the rotation frequency exceeds critical value (Kartashov, Malomed, and Torner [2007]). He, Malomed, and Wang [2007] showed that rotary motion of a two-dimensional soliton trapped in a Bessel lattice can be controlled by applying a finite-time push to the lattice, due to the transfer of linear momentum of the lattice to the orbital soliton momentum. Solitons in discrete rotating lattices were considered by Cuevas, Malomed, and Kevrekidis [2007].

## 7.2 Multipole and vortex solitons in azimuthally modulated Bessel lattices

Holographic techniques might be also used to produce higher-order azimuthally modulated beams and lattices. Lattices induced by azimuthally modulated Bessel beams of order $n$ have a functional shape: $R(\eta, \zeta) = J_n^2[(2b_{\mathrm{lin}})^{1/2}r]\cos^2(n\phi)$. The local lattice maxima situated closer to the lattice center are more pronounced than others, and form a ring of $2n$ guiding channels where solitons can be located. Such lattices support both individual solitons in any of the guiding channels and multipole solitons arranged into necklaces (Kartashov, Egorov, Vysloukh, Torner [2004]).

Complex soliton configurations in focusing media may be stable when the field changes its sign between neighboring channels. Thus, necklaces consisting of $2n$ out-of-phase spots may form in the guiding ring of $n$-th order. They are completely stable for high enough energy flow levels (see Fig. 14 for representative examples of necklace profiles). Another intriguing opportunity afforded by azimuthally modulated Bessel lattices is that a single soliton initially located in one of the guiding sites of lattice ring and launched tangentially to the guiding ring starts traveling along the consecutive guiding sites, so that it can even return into input site. Since solitons have to overcome a potential barrier when passing between sites, they radiate small fraction of energy and can be trapped in different sites, thus realizing azimuthal switching that can be controlled by the input soliton angle and energy flow (Fig. 14).

Localization of light in modulated Bessel optical lattices was observed experimentally by Fisher, Neshev, Lopez-Aguayo, Desyatnikov, Sukhorukov, Krolikowski, Kivshar [2006]. In the experiment, higher-order lattices were generated in photorefractive crystals by employing a phase-imprinting technique. The result of linear diffraction of low power beams was shown to depend crucially on specific location of the input excitation. Thus, diffraction was observed to be strongest for the excitation in the origin of the lattice and almost sup-pressed for excitation of a single lattice site. Nonlinearity resulted in strong light confinement in one of the sites of guiding lattice ring.

As it was discussed above, vortex solitons may be stabilized by imprinting of optical lattices in the focusing medium. However, the very refractive index modulation causing stabilization of vortex solitons simultaneously imposes certain restrictions on the possible topological charges of vortices dictated by the finite order of allowed discrete rotations (Ferrando, Zacares, Garcia-March [2005]). A corollary of such result is that the maximum charge of stable symmetric vortex in two-dimensional square lattices is equal to 1. Azimuthally modulated Bessel lattices offer a wealth of new opportunities because the order of rotational symmetry in such lattices may be higher than 4, in contrast to square lattices, a property that has direct implications in the possible topological charges of symmetric vortex solitons in Bessel lattices. By using general group theory arguments Kartashov, Ferrando, Egorov, Torner [2005] have shown that Bessel optical lattices may support vortex solitons with charges higher than one and that allowed values of vortex charge $m$ in a lattice of order $n$ is dictated by the charge rule $0 < m \le n - 1$. A detailed stability analysis have shown that vortex solitons in Bessel lattice may be stable only if the topological charge satisfies the condition $n/2 < m \le n - 1$, with the exception for $n = 2$, when the only existing vortex with $m = 1$ may be stable.

## 7.3    Soliton wires and networks

Arrays of lowest order Bessel beams might be used to produce reconfigurable two-dimensional networks in Kerr-type nonlinear media. The possibility of blocking and routing solitons in two-dimensional networks was suggested by Christodoulides, Eugenieva [2001a] and [2001b]; Eugenieva, Efremidis, Christodoulides [2001]. It was predicted that solitons can be navigated in two-dimensional networks and that this can be accomplished via vectorial interactions between two classes of solitons: broad and highly mobile signal solitons and powerful blockers. Interactions with signal and blocker beams combined with possibility for solitons to travel only along network wires might be used for implementation of a rich variety of blocking, routing, and logic operations. Solitons in such networks may propagate along sharp bends with angles exceeding 90 degrees, while engineering of the corner site of the bend allows to minimize bending losses.

While technological fabrication of such networks remain technically challenging, they could be induced optically with arrays of Bessel beams in the suitable nonlinear material. Solitons launched into such networks can travel along the complex predetermined paths (such as lines, corners, circular chains) almost without radiation losses (Xu, Kartashov, Torner [2005b]), while the whole network can be easily reconfigured just by blocking selected Bessel beams in network-inducing array. Several mutually incoherent Bessel beams can induce reconfigurable directional couplers and X-junctions (Xu, Kartashov, Vysloukh, Torner [2005]).

### 7.4    Mathieu and parabolic optical lattices

As it was mentioned above, Bessel beams are associated with the fundamental nondiffracting solutions of wave equation in circular cylindrical coordinates. Another important class of nondiffracting beams, so-called Mathieu beams, stem from the elliptical cylindrical coordinate system (Gutierrez-Vega, Iturbe-Castillo, Ramirez, Tepichin, Rodrigues-Dagnino, Chavez-Cerda, New [2001]). The lattices induced by Mathieu beams are particularly interesting because they afford smooth topological transformation of radially symmetric Bessel lattice into quasi-one-dimensional periodic lattice with modification of so-called interfocal parameter, thereby connecting two classes of optical lattices widely studied in the literature. The transformation of lattice topology finds its manifestation in important change of the shape and properties of ground-state and dipole-mode solitons. In particular, solitons are strongly pinned by almost radially symmetric lattices with small interfocal parameter and can hardly jump into neighbouring lattice rings, while in quasi-1D lattices with large interfocal parameters even small input tilts cause considerable transverse displacements of ground-state and dipole-mode solitons (Kartashov, Egorov, Vysloukh, Torner [2006]). The stability of dipole-mode solitons is determined by the very shape of the Mathieu lattice itself. Transformation of the lattice into a quasi-1D periodic structure destabilizes dipole solitons in the entire domain of their existence.

The unique properties exhibited by solitons in lattices produced by parabolic nondiffracting beams were introduce recently (Kartashov, Vysloukh, Torner [2008]). Parabolic lattices exhibit a nonzero curvature of their channels that results in asymmetric shapes of higher-order solitons. It was predicted that despite such symmetry breaking, complex higher-order states can be stable, and it was shown that the specific topology of parabolic lattices affords oscillatory-type soliton motion.

## §8.    Three-dimensional lattice solitons

In this section we briefly refer to the recent advances in the search for three-dimensional lattice solitons. Such solitons may form in 3D Bose-Einstein condensates loaded in optical lattices (for a recent review see Morsch, Oberthaler [2006]) as well as in potentially suitable nonlinear optical materials with a spatial refractive index modulation, when the full spatiotemporal dynamics is considered (for a recent review on spatiotemporal solitons, or light bullets, see Malomed, Mihalache, Wise, Torner [2005]). Notice that in the absence of guiding structures, 3D solitons in pure Kerr media experience strong collapse (for a review, see Berge [1998]). Under appropriate conditions optical lattices may completely stabilize ground-state and higher order 3D solitons, even in the case when dimensionality of the lattice is lower than that of the soliton.

The dynamics of 3D excitations in 3D or lower-dimensional optical lattices is described by the nonlinear Schrodinger equation that in the case of a focusing Kerr nonlinearity takes the form:

$$i\frac{\partial q}{\partial \xi} = -\frac{1}{2}\left(\frac{\partial^2 q}{\partial \eta^2} + \frac{\partial^2 q}{\partial \zeta^2} + \frac{\partial^2 q}{\partial \tau^2}\right) - q|q|^2 - pR(\eta, \zeta, \tau)q, \tag{8.1}$$

where for matter waves $\tau$ stands for third spatial coordinate and $\xi$ stands for evolution time, while for the spatiotemporal wave-packets $\tau$ is the time (we assume anomalous group-velocity dispersion) and $\xi$ is the propagation distance. The function $R(\eta, \zeta, \tau)$ describes profile of the lattice that can be of any dimensionality (up to 3D) in the case of matter waves and 2D or 1D for spatiotemporal solitons.

Three-dimensional periodic lattices with focusing nonlinearity can stabilize fundamental 3D solitons (Baizakov, Malomed, Salerno [2003]; Mihalache, Mazilu, Lederer, Malomed, Crasovan, Kartashov, Torner [2005]). The energy of 3D solitons in optical lattices always diverges at the cutoff and monotonically decreases as $b \to \infty$. However, a limited stability interval, where $dU/db > 0$, appears when the lattice depth exceeds a critical value. The width of the stability domain as a function of $b$ and $U$ increases with growing lattice depth. Illustrative profiles of 3D solitons in the 3D lattice $R = \cos(4\eta) + \cos(4\zeta) + \cos(4\tau)$ are depicted in Figs. 15(a)-(c), where transformation of broad unstable soliton covering many lattice sites into stable localized soliton with increase of peak amplitude is clearly visible. Importantly, the Hamiltonian versus energy diagram for 3D lattice solitons features two cuspidal points, resulting in a typical swallowtail loop. Although this loop is one of the generic patterns known in the catastrophe theory, it rarely occurs in physical models.

Repulsive Bose-Einstein condensate confined in a three-dimensional periodic optical lattice supports gap solitons and gap vortex states which are spatially localized in all three dimensions and possess nontrivial particle flow. Both on-site and off-site planar vortices with unit charge that are confined in third dimension by Bragg reflection were encountered in 3D lattice (Alexander, Ostrovskaya, Sukhorukov, Kivshar [2005]). Such planar states can be stable over a wide range of their existence region. Besides planar vortices with axes corresponding to X symmetry point in the Brillouin zone, solitons with other symmetry axes become possible in 3D lattices. 3D vortices in discrete systems were studied by Kevrekidis, Malomed, Frantzeskakis, Carretero-Gonzalez [2004]; Carretero-Gonzalez, Kevrekidis, Malomed, Frantzeskakis [2005]. In such systems stable vortices with $m = 1$ and 3 were encountered, while vortices with $m = 2$ are unstable and may spontaneously rearrange themselves into charge-3 vortices. Multicomponent discrete systems allow for existence of complex stable composite states, such as states consisting of two vortices whose orientations are perpendicular to each other.

Low-dimensional lattices may also support a variety of stable soliton states. Thus, quasi-2D lattices stabilize 3D solitons, while quasi-1D lattices may support stable 2D solitons in focusing cubic media (Baizakov, Malomed, Salerno [2004]; Mihalache, Mazilu, Lederer, Kartashov, Crasovan, Torner [2004]). Since lattice is uniform in one direction the solitons can freely move in that direction and experience head-on or tangential collisions, which may be almost elastic (as in the case of head-on collision of out-of-phase solitons) or may result in soliton fusion (for slow tangential collisions). Solitons in such lattices typically acquire strongly asymmetric shapes since they are better localized in dimension where lattice is imprinted than in uniform dimension. Notice that two cuspidal points in Hamiltonian versus energy diagram resulting in a swallowtail loop also appear for 3D solitons in quasi-2D lattices. Such solitons feature limited stability interval in $b$ or $U$ when lattice depth exceeds critical level, which is much higher than the critical depth for soliton stabilization in fully-3D lattice.

Cylindrical Bessel lattices imprinted in Kerr self-focusing media also support stable 3D solitons (Mihalache, Mazilu, Lederer, Malomed, Kartashov, Crasovan, Torner [2005]). In $J_0$ Bessel lattices, the fundamental solitons might be stable within one or even two intervals of their energy depending on the depth of the lattice. In the latter case the Hamiltonian versus energy diagram has a swallowtail shape with three cuspidal points. Stability properties for solitons in Bessel lattices are dictated to a large extent by the actual lattice shape and its asymptotical behavior at $r \to \infty$. Thus, $J_0^2$ lattices allow for single soliton stability domain only. While high-amplitude solitons are mostly confined in the central core of Bessel lattices, their low-amplitude counterparts acquire clearly pronounced multi-ring structure (see Figs. 15(d)-15(f)). Recently, the basic properties of three-dimensional vortex solitons in quasi-two dimensional lattices have been analyzed by Leblond, Malomed, and Mihalache [2007].

Quasi-1D lattices cannot stabilize 3D solitons in media with focusing Kerr nonlinearity (Baizakov, Malomed, Salerno [2004]). However, it was predicted recently that combination of quasi-1D lattice with Feshbach resonance management of scattering length in 3D Bose-Einstein condensates result in existence of stable 3D breather solitons (Matuszewski, Infeld, Malomed, Trippenbach [2005]). Such stable solitons may exist only if the average value of the nonlinear coefficient and the lattice strength exceed certain minimum values. Note that both 1D and 2D solitons in fully-dimensional optical lattices with nonlinearity management have been studied too (Gubeskys, Malomed, Merhasin [2005]). It was found, for example, that the energy thresholds required for existence of all types of two-dimensional solitons in dynamical systems is essentially higher than those in static systems. A detailed description of the soliton properties in systems with nonlinearity management is beyond the scope of this review. We refer to Malomed [2005] for a comprehensive description of the relevant background and methods.

## §9.   Nonlinear lattices and soliton arrays

One of the key ingredients required for the successful generation of lattice solitons is the robustness of the periodic optical pattern that effectively modulates the refractive index of the nonlinear medium, thereby creating a shallow grating whose depth is comparable with the nonlinear contribution to the refractive index produced by the soliton beam. In all previous sections the robustness of the optically induced lattice is achieved because the lattice propagates almost in the linear regime, so that self-action of the lattice and cross-action from the co-propagating soliton are negligible. In photorefractive media this can be achieved by selecting proper (ordinary) polarization for lattice waves (Fleischer, Segev, Efremidis, Christodoulides [2003]). At the same time, extraordinarily polarized lattices would experience strong self-action and may interact with soliton beams. The elucidation of conditions of robust propagation of such patterns is of interest, since it allows extension of the concept of optically induced gratings beyond the limit of weak material nonlinearities. In this section we briefly describe the properties of the nonlinear optical lattices and techniques that can be used for their stabilization.

## 9.1  Soliton arrays and pixels

A simple example of periodic nonlinear pattern is provided by an array of well-localized solitons. Such arrays can be created, for instance, by launching a spatially modulated incoherent beam in a focusing medium (Chen, McCarthy [2002]). The array of solitons generated in such way can be robust, provided that the coherence of the beam and the nonlinearity strength are not too high. If the coherence is too high, the array tends to break up into a disordered pattern rather than into ordered soliton structures. Thus, the incoherence plays a crucial role for stabilization of periodic patterns since it reduces the interaction forces between neighboring soliton pixels in the array.

Soliton arrays imprinted in photorefractive crystals with partially incoherent light might be used to guide probe beams at other wavelengths or for transmission of images. Such arrays can be created dynamically by seeding spatial noise onto a uniform partially spatially incoherent beam due to the development of the induced modulational instability. In strongly anisotropic photorefractive media, the dimensionality of the emerging pattern depends on the strength of the nonlinearity (Klinger, Martin, Chen [2001]). Large soliton arrays were created with coherent light too (Petter, Schroder, Träger, Denz [2003]). It was demonstrated that the attractive forces acting between in-phase solitons in such arrays might be exploited for controlled fusion of several solitons in array caused by additional beams launched between array channels. However, such attractive forces simultaneously put restrictions on the minimal separation between in-phase solitons in the array. Decreasing separation results in increase of interaction forces between neighboring spots and enhances the probability of fast amplification of asymmetries introduced by input noise or slight deviations of shapes of separate spots, that could lead to decay of the whole array.

The stability of soliton array can be substantially improved by utilizing phase-dependent interactions between soliton rows (Petrovic, Träger, Strinic, Belic, Schroder, Denz [2003]). In particular, changing the phase difference between neighboring rows of array by $\pi$ makes the interactions between them repulsive, prevents solitons in different rows from fusion, and greatly enhances stability of the entire array. The technique of phase engineering allows achieving a much denser soliton packing rate.

## 9.2    Nonlinear periodic lattices

The properties and stability of truly infinite two-dimensional stationary periodic waves in nonlinear media were analyzed by Kartashov, Vysloukh, Torner [2003]. Such patterns might be termed *cnoidal waves,* by analogy with their one-dimensional counterparts. In saturable nonlinear media such waves transform into harmonic patterns (akin to linear periodic lattices discussed in sections 5 and 6) in low- and high-power limits, while at intermediate power levels they describe arrays of localized bright spots in focusing media and arrays of dark solitons in defocusing media. Different cnoidal waves were found that can be divided into classes (3 in focusing media and 1 in defocusing media) depending on their shape (see Fig. 16).

A stability analysis of such periodic patterns has shown that all of them are unstable from a rigorous mathematical point of view, but practically the instability growth rates can be very small in two limiting cases of relatively low and high energy flows, indicating that nonlinearity saturation play strong stabilizing action for periodic patterns. Importantly, stability of patterns comprising out-of-phase spots and having chessboard phase structure (like cn-cn or sn-sn waves of Fig. 16) is substantially enhanced in comparison with that for patterns built of in-phase spots. Two-dimensional nonlinear lattices with chessboard phase structure were demonstrated in highly anisotropic photorefractive media (Desyatnikov, Neshev, Kivshar, Sagemerten, Träger, Jagers, Denz, Kartashov [2005]). The propagation of ra-diation in such medium under the steady-state conditions is described by the system of equations for dimensionless light field amplitude $q$ and electrostatic potential $\phi$ of the optically induced space-charge field $\mathbf{E}_{\mathrm{sc}} = \nabla U_{\mathrm{sc}}$:

$$i\frac{\partial q}{\partial \xi} = -\frac{1}{2}\left(\frac{\partial^2 q}{\partial \eta^2} + \frac{\partial^2 q}{\partial \zeta^2}\right) + \sigma q \frac{\partial \phi}{\partial \eta},$$

$$\frac{\partial^2 \phi}{\partial \eta^2} + \frac{\partial^2 \phi}{\partial \zeta^2} + \left(E + \frac{\partial \phi}{\partial \eta}\right)\frac{\partial}{\partial \eta}\ln(1 + S|q|^2) + \frac{\partial \phi}{\partial \zeta}\frac{\partial}{\partial \zeta}\ln(1 + S|q|^2) = 0, \tag{9.1}$$

where $q = AI_0^{-1/2}$, $\phi = (1/2)k^2 n^2 r_0 |r_{\mathrm{eff}}| U_{\mathrm{sc}}$, $I_0$ is the input intensity, $\sigma = \mathrm{sgn}(r_{\mathrm{eff}})$, $r_{\mathrm{eff}}$ is the effective electro-optic coefficient involved, $E = (1/2)k^2 n^2 r_0^2 E_0 |r_{\mathrm{eff}}|$ is the dimensionless static electric field applied to the crystal, $k$ is the wavenumber, $r_0$ is the beam width, $S = I_0/(I_{\mathrm{dark}} + I_{\mathrm{bg}})$ is the saturation parameter. It was found that due to anisotropy of photorefractive response refractive index modulation induced by periodic lattice is also highly anisotropic and nonlocal, and it depends on the lattice orientation relative to the crystal axis. Moreover, stability properties of stationary periodic waves in such media are also strongly affected by the pattern orientation: square patterns parallel to the crystal's c axis are less robust than diamond patterns oriented diagonally.

Besides lattices with the simplest chessboard phase structure a variety of self-trapped periodic patterns was demonstrated by Desyatnikov, Sagemerten, Fisher, Terhalle, Träger, Neshev, Dreischuh, Denz, Krolikowski, Kivshar [2006], including triangular lattices produced by interference of six plane waves (Fig. 17) and vortex lattices produced by waves with nested arrays of vortex-type phase dislocations. A detailed analysis of two-dimensional nonlinear self-trapped photonic lattices in anisotropic photorefractive media was carried out by Terhalle, Träger, Tang, Imbrock, and Denz [2006].

Nonlinear periodic structures may interact with localized soliton beams. Localized beams cause strong deformations of periodic patterns and under appropriate conditions they can form composite states (Desyatnikov, Ostrovskaya, Kivshar, Denz [2003]). Thus, periodic nonlinear lattice propagating in defocusing medium localizes other component in the form of a stable gap soliton. A variety of stationary composite lattice-soliton states of vector Kerr nonlinear Schrodinger equation were found via the Darboux transformation technique by Shin [2004]. In focusing media all such composite states were found to be unstable due to instability of periodic components, while the only stable state was found in defocusing medium.

Experimentally, one-dimensional composite gap solitons supported by nonlinear lattices were observed in $LiNbO_3$ crystals possessing saturable defocusing nonlinearity. Such states form when a Gaussian beam is launched at a Bragg angle into nonlinear lattice (Song, Liu, Guo, Liu, Zhu, Gao [2006]). In contrast to the case of fixed lattices transverse soliton mobility can be greatly enhanced in nonlinear lattices that experience deformations at soliton locations due to cross-modulation coupling (Sukhorukov [2006]).

Interaction of two-dimensional localized solitons with partially coherent nonlinear lattices was studied experimentally by Martin, Eugenieva, Chen, Christodoulides [2004]. Such lattices might be produced with amplitude masks that modulate otherwise uniform incoherent beams, while changing the degree of incoherence allows to control stability of periodic pattern. When the intensity of the soliton is comparable with that of the lattice and it is launched at small angle relative to lattice propagation direction, the lattice dislocation is created due to soliton dragging. In this case transverse soliton shift is much smaller in the presence of the lattice due to strong soliton-lattice interaction.

Polaron-like structure may form in partially coherent nonlinear lattice with small enough spacing between its spots. In such structures soliton drags toward it some of lattice sites while pushing away the other. Interaction between vortex beams and nonlinear partially coherent lattice may result in lattice twisting driven by the angular momentum carried by the vortex beam, with the direction of twisting being opposite for vortices with opposite charges (Chen, Martin, Bezryadina, Neshev, Kivshar, Christodoulides [2005]). The formation of two-dimensional composite solitons upon interaction of mutually incoherent lattice and stripe beams was observed by Neshev, Kivshar, Martin, Chen [2004]. Such interaction results in decrease of spacing between lattice rows in the interaction region while moving stripes cause lattice compression and deformation.

## §10.  Defect modes and random lattices

In this section we summarize extensive theoretical and experimental activity of different research groups in the field of so called defect modes. Some structural imperfections are inevitable during fabrication of waveguide array as well as upon optical lattice induction, so their influence on the properties of solitons should be carefully studied. Moreover, specially designed defects could even be applied for controllable filtering, switching, and steering of optical beams in lattices.

## 10.1 Defect modes in waveguide arrays and optically-induced lattices

The propagation of discrete solitons in a waveguide array in the presence of localized coupling constant perturbations was studied by Krolikowski and Kivshar [1996]. Some potential schemes for controllable soliton switching and steering utilizing array defects were discussed. Later, suppression of defect-induced guiding due to focusing nonlinearity in Al-GaAs waveguide array was demonstrated experimentally (Peschel, Morandotti, Aitchison, Eisenberg, and Silberberg [1999]). The existence and stability of nonlinear localized waves in 1D periodic medium described by the Kronig-Penney model with a nonlinear defect was reported by Sukhorukov and Kivshar [2001]. A novel type of stable nonlinear band-gap localized state has been found. The existence and stability of bright, dark, and twisted spatially localized modes in arrays of thin-film nonlinear waveguides that can be viewed as multiple point-like nonlinear defects in otherwise linear medium (Dirac-comb nonlinear lattice) was presented by the same authors (Sukhorukov, Kivshar [2002b]). An overview of the properties of nonlinear guided waves in such structures can be found in the paper by Sukhorukov and Kivshar [2002c], where authors discuss similarities and differences between discrete and continuous models.

Recently, Kominis [2006] and Kominis, Hizanidis [2006] applied phase-space method for construction of analytical bright and dark soliton solutions of the nonlinear Kronig-Penney model describing periodic photonic structure with alternating linear and nonlinear layers. The existence of stable nonlinear localized defect modes near the band edge of 2D reduced-symmetry photonic crystal with Kerr nonlinearity was predicted by Mingaleev and Kivshar [2001]. The physical mechanism resulting in the mode stabilization was revealed. A new type of waveguide lattice that relies on the effect of bandgap guidance in the regions between waveguide defects was proposed by Efremidis and Hizanidis [2005]. It was shown that spatial solitons that emerge from different spectral gaps of periodic binary waveguide array can be selectively reflected or transmitted through an engineered defect, which acts as a low- or high-pass filter (Sukhorukov and Kivshar [2005]). Defect modes in 1D optical lat-tice were recently discussed by Brazhnyi, Konotop, and Perez-Garcia [2006a] in the context of matter waves transport in optical lattices. Such modes supported by the localized defects in otherwise periodic lattice can be accurately described by an expansion over Wannier functions, whose envelope is governed by the coupled nonlinear Schrodinger equations with a $\delta$-impurity (Brazhnyi, Konotop, Perez-Garcia [2006b]).

The interaction of moving discrete solitons with defects in 1D fabricated arrays of semiconductor waveguides was investigated experimentally by Morandotti, Eisenberg, Mandelik, Silberberg, Modotto, Sorel, Stanley, Aitchison [2003]. Important changes of the output soliton positions (sudden switch across a narrow repulsive defect) were observed for relatively small changes of input conditions. The formation of defect modes in polymer waveguide arrays was investigated both theoretically and experimentally by Trompeter, Peschel, Pertsch, Lederer, Streppel, Michaelis, and Brauer [2003]. It was shown that variation in the defect strength (or effective index) is accompanied by the modification of character and number of modes bound to the defect. Although the symmetric defect waveguide becomes multimode with increase of effective index it does not support antisymmetric modes.

The dynamical properties of 1D discrete NLSE with arbitrary distribution of on-site defects were considered theoretically by Trombettoni, Smerzi, and Bishop [2003]. Several possible regimes of beam propagation including its complete reflection, oscillations around fixed points and formation of self-trapped states have been predicted. The trapping of moving discrete soliton by linear and nonlinear impurities in discrete arrays was recently discussed by Morales-Molina and Vicencio [2006].

Linear defect modes can also exist in photoinduced lattices with both positive (locally increased refractive index) and negative (locally decreased refractive index) defects. Guided modes supported by relatively extended positive and negative defects in hexagonal optical lattices and far-field power spectra of probe beams propagating in such lattices that clearly reveal the physical mechanisms leading to localization of linear modes (total internal or Bragg reflection) were presented by Bartal, Cohen, Buljan, Fleischer, Manela, Segev [2005]. The band-gap guidance of light in optically induced 2D lattice with single-site negative defect was observed by Makasyuk, Chen, Yang [2006]. Defect modes supported by such lattices feature long tails in the directions of the lattice principal axes as shown in Fig. 18.

In linear case the strongest defect mode confinement appears when the lattice intensity at the defect site is nonzero rather than zero (Fedele, Yang, and Chen [2005]; Wang, Yang, and Chen [2007]). It was shown that in nonlinear regime defect solitons bifurcate from every infinitesimal linear defect mode (Yang, Chen [2006]). At high powers in the medium with focusing nonlinearity defect soliton modes become unstable in the case of positive defects, but may remain stable in the case of negative defect.

Defect solitons in optically induced one-dimensional photonic lattices in $LiNbO_3$:Fe crystals were observed experimentally by Qi, Liu, Guo, Lu, Liu, Zhou, and Li [2007]. Wave and defect dynamics in 2D nonlinear photonic quasicrystals was studied experimentally by Freedman, Bartal, Segev, Lifshitz, Christodoulides, and Fleischer [2006]. The main properties of solitons supported by defects embedded in superlattices have been analyzed by Chen, He, and Wang [2007].

## 10.2  Anderson localization

The concept of Anderson localization was originally introduced in the field of condensed matter physics for the phenomenon of disorder-induced metal-insulator transition in linear electronic systems. Anderson localization refers to the situation where electron, when released inside a random medium, may stay close to the initial point (Anderson [1978]). The mechanism behind this phenomenon has been attributed to multiple scattering of electron by the random potential, a feature of the wave nature of electrons. This concept may also be applied to the classical linear wave systems (Maynard [2001]).

Different linear regimes of light localization in disordered photonic crystal were considered theoretically and experimentally by Vlasov, Kaliteevski, and Nikolaev [1999]. In particular, it was shown that Bloch states are disrupted only when local fluctuations of the band-edge frequency (caused by randomization of refractive index profile) become as large as the band-gap width. Band theory of light localization in one-dimensional linear disordered systems was presented by Vinogradov and Merzlikin [2004]. It was detected that Bragg reflection is responsible not only for appearance of band-gaps but also for Anderson localization of light in one-dimensional case.

Recently Schwartz, Bartal, Fishman, and Segev [2007] reported on the experimental observation of Anderson localization in a perturbed periodic potentials. In the setup used by authors the transverse localization of light was caused by random fluctuations on a two-dimensional photonic lattice imprinted in a photorefractive crystal. Authors demonstrated how ballistic transport becomes diffusive in the presence of disorder, and that crossover to Anderson localization occurs at high levels of disorder. In this experiment the controllable level of disorder was achieved by combining random and regular lattices, while optical induction technique allowed easy creation of required ensemble of lattice realizations. Anderson localization in one-dimensional disordered waveguide arrays has been observed very recently by Lahini, Avidan, Pozzi, Sorel, Morandotti, Christodoulides, and Silberberg [2008].

From the viewpoint of fundamental nonlinear physics the challenging problem is the exploration of nonlinear analogues of Anderson localization. The interplay between disorder and nonlinearity was studied in systems described by one-dimensional nonlinear Schrödinger equation with random-point impurities (see, e.g., Kivshar, Gredeskul, Sanchez, Vazquez [1990]; Hopkins, Keat, Meegan, Zhang, Maynard [1996]) as well as in 1D (Kevrekidis, Kivshar, Kovalev [2003]) and 2D (Pertsch, Peschel, Kobelke, Schuster, Bartelt, Nolte, Tünnermann, Lederer [2004]) discrete waveguide arrays. Anderson-type localization of moving spatial solitons in optical lattices with random frequency modulation was studied by Kartashov, Vysloukh [2005]. In this case dramatic enhancement of soliton trapping probability on lattice inhomogeneities was detected with increase of frequency fluctuation level. The localization process is strongly sensitive to lattice depth since in shallow lattices moving solitons experience random refraction or/and multiple scattering, in contrast to relatively deep lattices, where solitons are typically immobilized in the vicinity of modulation frequency local minima. It is worth noticing that analogous phenomena were encountered in trapped BECs (Greiner, Mandel, Esslinger, Hänsch, Bloch [2002]; Abdullaev, Garnier [2005]; Kuhn, Miniatura, Delande, Sigwarth, Muller [2005]; Schulte, Drenkelforth, Kruse, Ertmer, Arlt, Sacha, Zakrzewski, Lewenstein [2005]; Gavish and Castin [2005]).

## 10.3  Soliton percolation

The combined effect of periodic and random potentials on the transmission of moving and formation of stable stationary excitations has been addressed in a number of studies and reviews (Bishop, Jimenez, Vazquez [1995]; Abdullaev, Bishop, Pnevmatikos, and Economou [1992]; Abdullaev [1994]). A universal feature of wave packet and particle dynamics in disordered media in different areas of physics is percolation (Grimmett [1999]); Shklovskii, Efros [1984]). Percolation occurs in different types of physical settings, including high-mobility electron systems, Josephson-junction arrays, two-dimensional GaAs structures near the metal-insulator transition, or charge transfer between superconductor and hoping insulator, to mention a few.

Recently the nonlinear optical analog of biased percolation was introduced (Kartashov, Vysloukh, Torner [2007c]). The phenomenon analyzed is the disorder-induced soliton transport in randomly modulated optical lattices with Kerr-type focusing nonlinearity in the presence of a linear variation of the refractive index in the transverse plane, thus generating a constant deflecting force for light beams entering the medium. When such a force is small enough, solitons in perfectly periodic lattices are trapped in the vicinity of the launching point due to the Peierls-Nabarro potential barriers, provided that the launching angle is smaller than a critical value. Under such conditions soliton transport is suppressed, and thus the lattice acts as a soliton insulator. However, random modulations of the lattice parameters makes soliton mobility possible again, with the key parameter determining the soliton current being the standard deviation of the phase/amplitude fluctuations. Kartashov, Vysloukh, Torner [2007c] uncovered that the soliton current in lattices with amplitude and phase fluctuations reaches its maximal value at intermediate disorder levels and that it drastically reduces in both, almost regular and strongly disordered lattices. This suggests the possibility of a disorder-induced transition between soliton insulator and soliton conductor lattice states.

## §11. Concluding remarks

To conclude, we would like to add some comments on two important fields, emerging from theoretical and experimental research in the area of lattice solitons, which fall out of the main stream of this review.

The first topic is the generalization of the lattice soliton concept to partially coherent or incoherent (such as white-light) optical fields. From a physical point of view, the diffraction spreading of a partially coherent light beam is determined by the transverse coherence length rather than by the beam radius. In shallow quasicontinuous lattices, this means that self-trapping and soliton formation would normally require higher power levels for incoherent radiation. In this case, a ratio between transverse coherence length and lattice period becomes crucial, since the soliton formation process is sensitive to the phase relations be-tween several neighboring lattice channels. Thus, the complex interplay between coherence, nonlinearity, and waveguiding becomes central.

A powerful theoretical approach to describe the propagation and self-focusing of partially spatially incoherent light beams in bulk photorefractive media was developed by Christodoulides, Coskun, Mitchell, and Segev [1997]; Mitchell, Segev, Coskun, and Christodoulides [1997]. It was followed by the experimental observation of self-trapping of white light in bulk media (Mitchell and Segev [1997]). Further theoretical progress was reported by Buljan, Segev, Soljacic, Efremidis, and Christodoulides [2003]. The existence of random phase solitons in nonlinear periodic lattices was predicted by Buljan, Cohen, Fleischer, Schwartz, Segev, Musslimani, Efremidis, and Christodoulides [2004]. Importantly, such solitons exist when the characteristic response time of the medium is much longer than the coherence time. The prediction mentioned above was confirmed experimentally (Cohen, Bartal, Buljan, Carmon, Fleischer, Segev, and Christodoulides [2005]). Lately the existence of lattice solitons made of incoherent white light, originating from an ordinary incandescent light bulb, was analyzed by Pezer, Buljan, Bartal, Segev, and Fleisher [2006]. Prospects of incoherent gap optical soliton formation in a self-defocusing nonlinear periodic medium was analyzed theoretically (Motzek, Sukhorukov, Kaiser, and Kivshar [2005]; Pezer, Buljan, Fleischer, Bartal, Cohen, and Segev [2005]), and further supported by the experimental observation (Bartal, Cohen, Manela, Segev, Fleischer, Pezer, and Buljan [2006]). The phenomenon of modulational instability of extended incoherent nonlinear waves waves in photonic lattices was addressed by Jablan, Buljan, Manela, Bartal, Segev [2007].

A second intriguing and practically important topic directly linked with lattice solitons is the physics of nonlinear surface wave formation at the interface of periodic structures. Surface waves were originally introduced by Igor E. Tamm in 1932 in the context of condensed-matter physics and have been intensively studied in different areas of science. However the progress in experimental investigation of nonlinear optical surface waves was severely limited by high power thresholds required for their formation. Recently it was predicted theoretically (Makris, Suntsov, Christodoulides, Stegeman and Hache [2005]) and confirmed experimentally (Suntsov, Makris, Christodoulides, Stegeman, Hache, Morandotti, Yang, Salamo, and Sorel [2006]) that surface waves formation at the very edge of one-dimensional waveguiding array is accessible for moderate power levels. Formation of gap surface solitons at the edge of defocusing lattice is also possible as it was shown theoretically (Kartashov, Vysloukh, and Torner [2006]) and confirmed experimentally in defocusing LiNbO$_3$ waveguiding arrays (Rosberg, Neshev, Krolikowski, Mitchell, Vicencio, Molina, and Kivshar [2006]; Smirnov, Stepic, Rüter, Kip, and Shandarov [2006]). Siviloglou, Makris, Iwanow, Schiek, Christodoulides, Stegeman, Min, and Sohler [2006] reported on experimental observation of discrete quadratic surface solitons in self-focusing and defocusing periodically poled lithium niobate waveguide arrays. Lattice interfaces imprinted in nonlocal nonlinear media support interesting families of multipole surface solitons (Kartashov, Torner, and Vysloukh [2006]).

Two-dimensional geometry offers even richer possibilities for surface waves formation (Kartashov and Torner [2006]; Kartashov, Vysloukh, Mihalache, and Torner [2006]; Kartashov, Vysloukh, and Torner [2006]; Makris, Hudock, Christodoulides, Stegeman, Manela, and Segev [2006]). Solitons residing at the lattice edges, in the corners, and on specially designed defects may serve here as illuminating examples. Experimental observations of two-dimensional surface waves were realized at the boundaries of a finite optically induced lattice (Wang, Bezryadina, Chen, Makris, Christodoulides, and Stegeman [2007]), and in laser-written waveguide arrays (Szameit, Kartashov, Dreisow, Pertsch, Nolte, Tünnermann, and Torner [2007]). Many problems remain open to explore the combination of surface solitons and optical lattices.

We thus conclude by stressing that we presented a concise review of the latest advances in optical soliton formation, manipulation, shaping, and control in optical lattices, with special emphasis on optically-induced lattices. Many concepts investigated in the context of optical lattices have implications and applications in several branches of physics, such as nonlinear optics, condensed-matter theory, and quantum mechanics. Optical soliton control in optical lattices offers a unique laboratory to investigate universal phenomena, and as such provides a unique exploration tool.

## §12.   Acknowledgements

We thank all our colleagues and collaborators for invaluable discussions over the last few years in conducting joint research reported here and in completing this review. Most especially we thank the members of our group, the co-authors of joint papers described here, G. Assanto, R. Carretero-Gonzalez, Z. Chen, D. Christodoulides, L. Crasovan, C. Denz, A. De-

syatnikov, A. Ferrando, D. Kip, Y. Kivshar, F. Lederer, M. Lewenstein, B. Malomed, D. Mazilu, D. Mihalache, M. Molina, D. Neshev, T. Pertsch, M. Segev, G. Stegeman, A. Szameit, Z. Xu, F. Ye for contributing their knowledge and figures included in this review. This work was produced with financial support by the Generalitat de Catalunya and by the Government of Spain through the Ramon-y-Cajal program and grant TEC2005-07815.
.

**Figure captions**

Figure 1.     (a) Dependencies of propagation constants of Bloch waves from different bands on transverse wavenumber $k$ at $\Omega = 4$ and $p = 3$. (b) Band-gap lattice structure at $\Omega = 4$. Bands are marked with gray color, gaps are shown white. Profiles of Bloch waves from first (c),(d) and second (e),(f) bands, corresponding to $k = 0$ (panels (c) and (e)) and $k = 2$ (panels (d) and (f)) at $\Omega = 4$ and $p = 3$.

Figure 2.     Typical profile of two-dimensional photonic lattice induced by the interference of four plane waves in highly anisotropic SBN:75 crystal. Each waveguide is approximately $7\ \mu$m in diameter, with an $11\ \mu$m spacing between nearest neighbors (Fleischer, Segev, Efremidis, Christodoulides [2003]).

Figure 3.     Lattices induced by the nondiffracting beams of different type: (a) zero-order radially symmetric Bessel lattice; (b) first-order radially symmetric Bessel lattice; (c) lattice produced by the azimuthally modulated Bessel beam of third order; odd Mathieu lattices with small (d) and large (e) interfocal parameters that closely resemble azimuthally modulated Bessel and quasi-1D periodic lattices, correspondingly; (f) lattice produced by the combination of odd and even parabolic beams.

Figure 4.     Odd (a), even (b), and twisted (c) solitons originating from semi-infinite gap of optical lattice with focusing nonlinearity at $b = 1.4$. Odd (d), even (e), and twisted (f) gap solitons originating from first finite gap of optical lattice with defocusing nonlinearity at $b = -0.5$. In all cases lattice depth $p = 4$.

Figure 5.     Experimental demonstration of odd and even spatial solitons. (a) Input beam and optical lattice (power $23\ \mu$W). (b) Output probe beam at low power ($2 \times 10^{-3}\ \mu$W). (c) Localized states ($87 \times 10^{-3}\ \mu$W). Left, propagation in the absence of the grating; middle, even excitation; right, odd excitation. Biasing

electric field 3600 V/cm (Neshev, Ostrovskaya, Kivshar, and Krolikowski [2003]).

Figure 6.    (a) Experimental and theoretical (inset) linear output in a straight lattice. (b) Shift of the nonlinear probe beam output versus the modulating beam power for $k_{3x} = 1.18k_{12x}$ and $\varphi/2\pi = 0.22$, where $k_{3x}$ and $k_{12x}$ stand for transverse wavenumbers of third and first two waves, $\varphi$ is the phase difference between third wave and other two waves. (c),(d) Experimental and theoretical (inset) nonlinear output in straight and modulated lattice for intensity of third wave $I_3 = 0$ and $I_3 = 4I_{12}$, respectively, and $k_{3x} = 1.18k_{12x}$. (e),(f) Numerical simulations of the longitudinal propagation (Rosberg, Garanovich, Sukhorukov, Neshev, Krolikowski, Kivshar [2006]).

Figure 7.    (Top) Cell for liquid-crystal waveguide array: $\Lambda$, array period, $d$, cell thickness, $V$, applied voltage, ITO, indium tin oxide electrodes. (Bottom) Experimental results with light propagation at 1064 nm wavelength: (a) one-dimensional diffraction at $V = 0$ V, (b) low-power discrete diffraction at $V = 1.2$ V and 20 mW input power, (c) discrete nematicon at $V = 1.2$ V and 35 mW input power. The residual coupling and transverse-longitudinal power fluctuations shown are due partly to the limits of the camera and partly to nonuniformities in the sample (Fratalocchi, Assanto, Brzdakiewicz, Karpierz [2004]).

Figure 8.    Experimental results presenting the propagation of a probe beam launched into a single waveguide at normal incidence. (a) Intensity structure of the probe beam at the exist face of the crystal, displaying discrete diffraction at low nonlinearity (200 V). (b) Intensity structure of a probe beam at the exit face of the crystal, displaying an on-axis lattice soliton at high nonlinearity (800 V). (c) Interferogram, showing constructive interference of peak and lobes between the soliton and a plane wave. (d) Reduction of probe intensity by a factor of 8, at the same voltage as (b) and (c), results in recovery of discrete diffraction pattern (Fleischer, Segev, Efremidis, Christodoulides [2003]).

Figure 9.     Experimental results of charge-4 vortex propagating without (top) and with
              (bottom) the 2D lattice for different applied electric fields. (a) input, (b) lin-
              ear diffraction, (c) output at a bias field of 90 V/mm, (d) output at
              240 V/mm, (e) discrete diffraction at 90 V/mm, (f)-(h) discrete trapping at
              240 V/mm. From (f) to (h), the vortex was launched into lattice from off-
              site, to on-site, and then back to off-site configurations. Distance of propaga-
              tion through the crystal is 20 mm; average distance between adjacent spots in
              the necklace is 48 $\mu$m (Yang, Makasyuk, Kevrekidis, Martin, Malomed,
              Frantzeskakis, Chen [2005]).

Figure 10.    Experimental observation of gap vortex lattice soliton. (a) Intensity distribu-
              tion of the input vortex-ring beam photographed (for size comparison) on the
              background of the optically induced lattice. (b) Output intensity distribution
              of a low intensity ring beam experiencing linear diffraction in the lattice. (c)
              Output intensity distribution of a high intensity ring beam, which has evolved
              into gap vortex soliton in the same lattice as in (b). (d)-(f) Phase structure of
              gap vortex lattice soliton. The phase information is obtained by interference
              with a weakly diverging Gaussian beam. (d) Phase distribution of the input
              vortex-ring beam. (e) Output phase distribution of a high intensity ring-beam
              that has evolved into gap vortex soliton (interference pattern between the
              soliton whose intensity is presented in (c) and a weakly diverging Gaussian
              beam). (f) Numerical validation of the phase information in (e) (Bartal,
              Manela, Cohen, Fleischer, Segev [2005]).

Figure 11.    (a) and (b) show lattices with single dislocations produced by the interference
              pattern of plane wave and wave with nested charge-1 vortex for various vor-
              tex orientations. Arrows show direction of motion for solitons placed into lat-
              tice channels in the vicinity of dislocation. Positive (c) and negative (d) topo-
              logical traps created with two vortices with charges $m = \pm 3$. Map of $F_\zeta$ force
              for solitons launched in the vicinity of positive (e) and negative (f) topological

traps produced by two vortices with charges $m = \pm 1$. Arrows show direction of soliton motion. In (c)-(f) trap length $\delta\zeta = \pi$ (Kartashov, Vysloukh, Torner [2005b]).

Figure 12.     (a) Interaction of out-of-phase solitons in the first ring of zero-order Bessel lattice. One of solitons is set into motion at the entrance of the medium by imposing the phase twist $\nu = 0.1$. (b) The same as in column (a) but for in-phase solitons. (c) Interaction of out-of-phase solitons located in first and second lattice rings. The soliton located in the first ring is set into motion by imposing the phase twist $\nu = 0.2$. All solitons correspond to the propagation constant $b = 5$ at $p = 5$. Soliton angular rotation directions are depicted by arrows (Kartashov, Vysloukh, Torner [2004c]).

Figure 13.     Soliton rotation in ring lattice. From (a) to (c), the probe soliton beam was aimed at the far right side of the first, second, and third lattice ring with the same tilting angle (about 0.4°) in the $y$ direction. From (d) to (f), the probe soliton beam was aimed at the same second lattice ring but with different tilting angles of 0°, 0.4°, and 0.6°, respectively (Wang, Chen, Kevrekidis [2006]).

Figure 14.     Stable soliton complex supported by third-order Bessel lattice at $b = 3$ and $p = 20$ (a) and by sixth-order Bessel lattice at $b = 5$ and $p = 40$ (b). Azimuthal switching of single tilted soliton to second (c) and fourth (d) channels of third-order Bessel lattice for tilt angles $\alpha_\zeta = 0.49$ and $\alpha_\zeta = 0.626$ at $p = 2$ and input energy flow $U_{in} = 8.26$. Panels (e) and (f) show switching to third and sixth channels of sixth-order lattice for $\alpha_\zeta = 0.8$ and $\alpha_\zeta = 0.93$ at $p = 5$ and input energy flow $U_{in} = 8.61$. Input and output intensity distributions are superimposed. Arrows show direction of soliton motion and $S_{in}, S_{out}$ denote input and output soliton positions (Kartashov, Egorov, Vysloukh, Torner [2004]).

Figure 15.     Isosurface plots of (a) unstable and (b),(c) stable 3D solitons in 3D periodic lattice. In (a) soliton energy $U = 2.4$ and its peak amplitude $A = 1.8$, in (b)

$U = 2.04$, $A = 2.2$, in (c) $U = 2.4$, $A = 3$. In (a)-(c) depth of periodic lattice $p = 3$. (d)-(f) Isosurface plots of stable 3D solitons supported by 2D Bessel lattice at (d) $p = 5$, $b = 0.11$, (e) $p = 5.5$, $b = 0.85$, and (f) $p = 6$, $b = 2$ (Mihalache, Mazilu, Lederer, Malomed, Crasovan, Kartashov, Torner [2005]; Mihalache, Mazilu, Lederer, Malomed, Kartashov, Crasovan, Torner [2005]).

Figure 16. Cn-cn (a) and sn-sn (b) periodic waves in saturable medium with $S = 0.1$. Plots in the first row show energy flow concentrated within one wave period $T = 2\pi$ versus propagation constant. Surface plots in each column show evolution of wave shape with increase of energy flow $U$. Column (a) corresponds to focusing nonlinearity, column (b) corresponds to defocusing nonlinearity (Kartashov, Vysloukh, Torner [2003]).

Figure 17. Theoretical (color) and experimental (grayscale) results for self-trapped triangular lattices created by interference of six plane waves, see far field images in the top panel. Two distinct orientations of the triangular lattices with respect to crystallographic $c$-axis are compared (top and bottom panels) as well as low (left) and high (right) saturation regimes. LW, lattice field (color) and intensity (grayscale) of the lattice wave. GW, calculated profiles of the refractive index (color) and measured intensity of the probe linear wave guided by the lattice (grayscale) (Desyatnikov, Sagemerten, Fisher, Terhalle, Träger, Neshev, Dreischuh, Denz, Krolikowski, Kivshar [2006]).

Figure 18. Top panel: Intensity pattern of a 2D optically induced lattice with a single-site negative defect at crystal input (a) and output (b), (c). The defect disappears in the linear regime (b) but can survive with weak nonlinearity (c) after propagating through a 20-mm-long crystal. Bottom panel: Input (a) and output (b), (c) of a probe beam showing band-gap guidance by the defect (c) and normal diffraction without the lattice (b) under the same bias condition (Makasyuk, Chen, Yang [2006]).

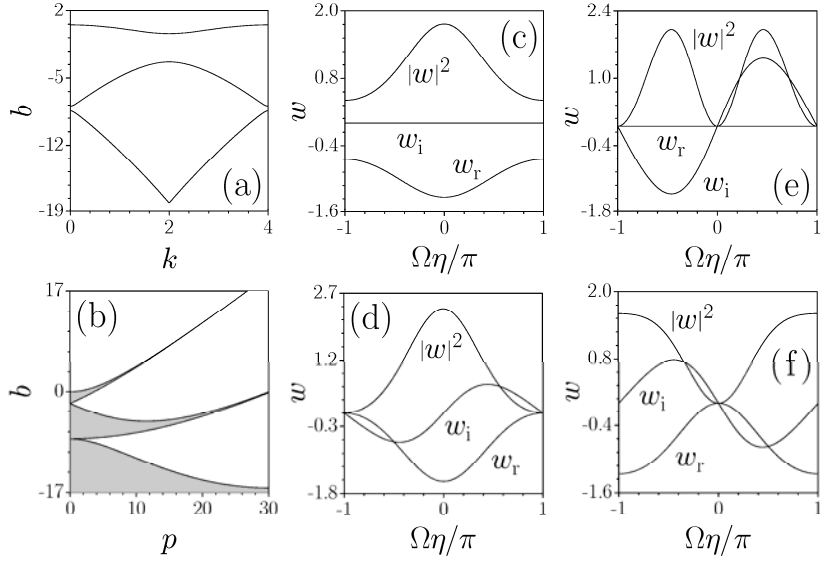

Figure 1. (a) Dependencies of propagation constants of Bloch waves from different bands on transverse wavenumber $k$ at $\Omega = 4$ and $p = 3$. (b) Band-gap lattice structure at $\Omega = 4$. Bands are marked with gray color, gaps are shown white. Profiles of Bloch waves from first (c),(d) and second (e),(f) bands, corresponding to $k = 0$ (panels (c) and (e)) and $k = 2$ (panels (d) and (f)) at $\Omega = 4$ and $p = 3$.

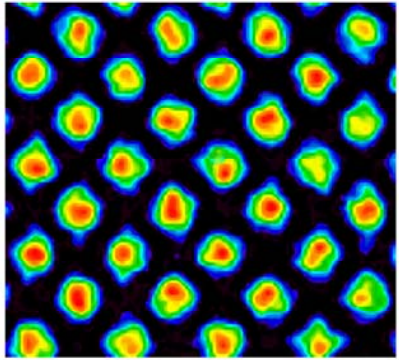

Figure 2.    Typical profile of two-dimensional photonic lattice induced by the inter-
ference of four plane waves in highly anisotropic SBN:75 crystal. Each
waveguide is approximately 7 $\mu$m in diameter, with an 11 $\mu$m spacing
between nearest neighbors (Fleischer, Segev, Efremidis, Christodoulides
[2003]).

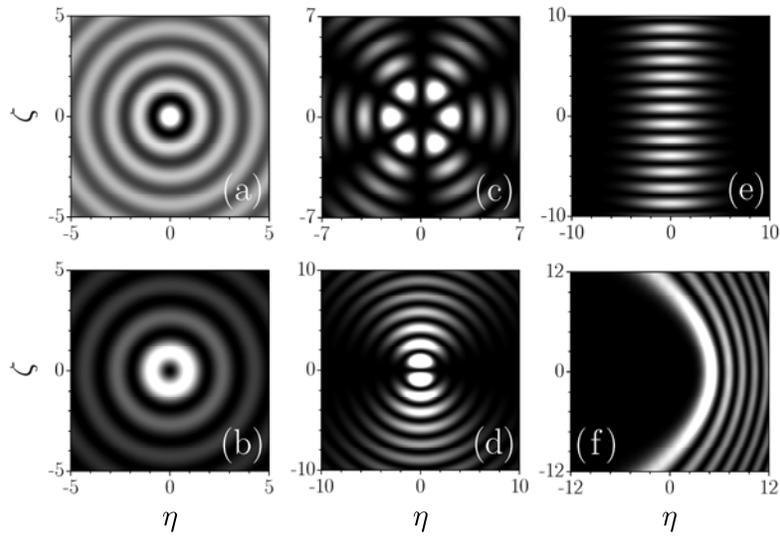

Figure 3. Lattices induced by the nondiffracting beams of different type: (a) zero-order radially symmetric Bessel lattice; (b) first-order radially symmetric Bessel lattice; (c) lattice produced by the azimuthally modulated Bessel beam of third order; odd Mathieu lattices with small (d) and large (e) interfocal parameters that closely resemble azimuthally modulated Bessel and quasi-1D periodic lattices, correspondingly; (f) lattice produced by the combination of odd and even parabolic beams.

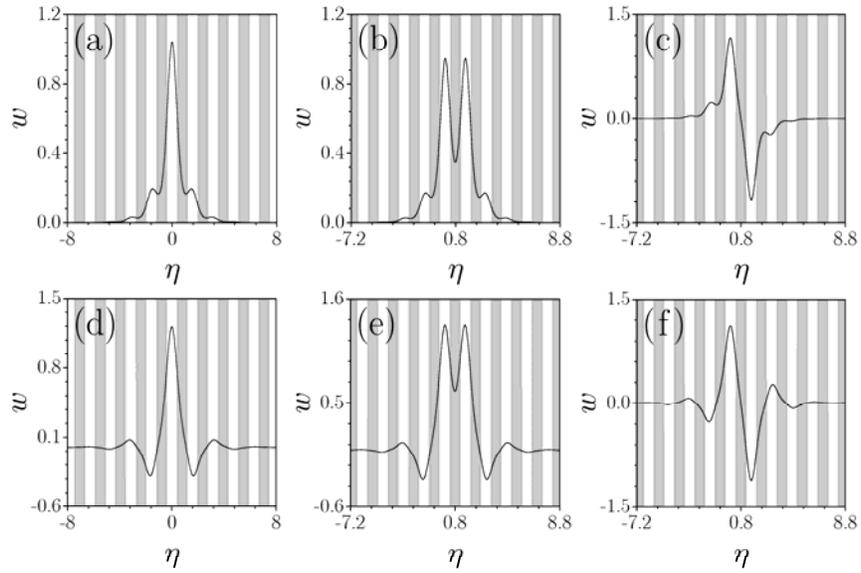

Figure 4. Odd (a), even (b), and twisted (c) solitons originating from semi-infinite gap of optical lattice with focusing nonlinearity at $b = 1.4$. Odd (d), even (e), and twisted (f) gap solitons originating from first finite gap of optical lattice with defocusing nonlinearity at $b = -0.5$. In all cases lattice depth $p = 4$.

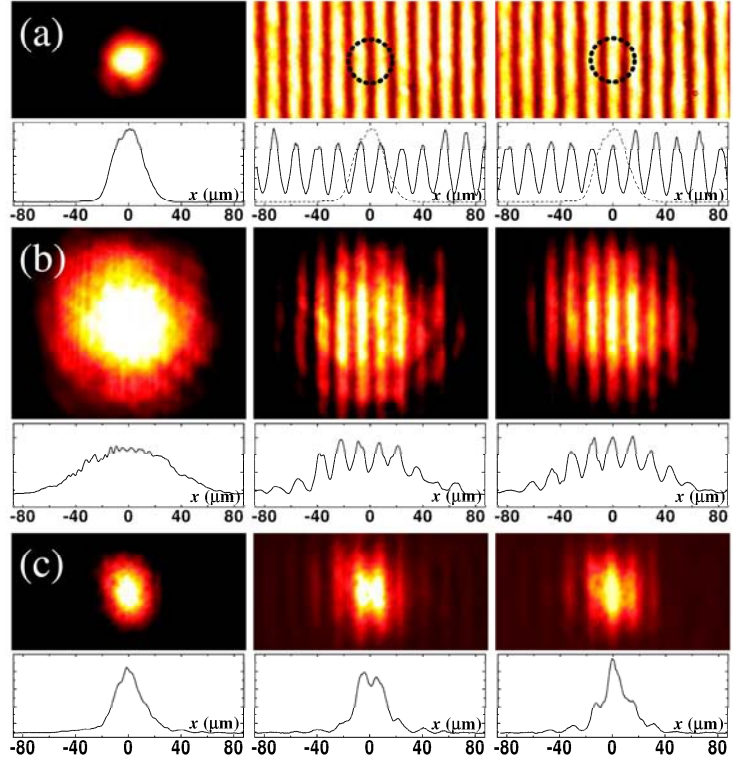

Figure 5.    Experimental demonstration of odd and even spatial solitons. (a) Input beam and optical lattice (power 23 $\mu$W). (b) Output probe beam at low power ($2 \times 10^{-3}$ $\mu$W). (c) Localized states ($87 \times 10^{-3}$ $\mu$W). Left, propagation in the absence of the grating; middle, even excitation; right, odd excitation. Biasing electric field 3600 V/cm (Neshev, Ostrovskaya, Kivshar, and Krolikowski [2003]).

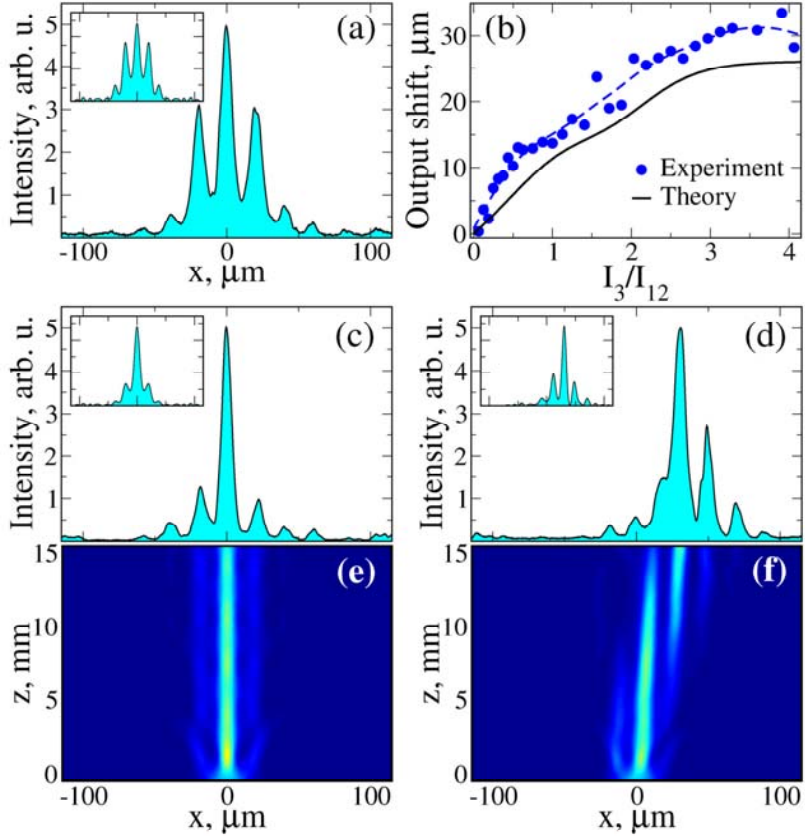

Figure 6.    (a) Experimental and theoretical (inset) linear output in a straight lattice. (b) Shift of the nonlinear probe beam output versus the modulating beam power for $k_{3x} = 1.18 k_{12x}$ and $\varphi/2\pi = 0.22$, where $k_{3x}$ and $k_{12x}$ stand for transverse wavenumbers of third and first two waves, $\varphi$ is the phase difference between third wave and other two waves. (c),(d) Experimental and theoretical (inset) nonlinear output in straight and modulated lattice for intensity of third wave $I_3 - 0$ and $I_3 - 4I_{12}$, respectively, and $k_{3x} = 1.18 k_{12x}$. (e),(f) Numerical simulations of the longitudinal propagation (Rosberg, Garanovich, Sukhorukov, Neshev, Krolikowski, Kivshar [2006]).

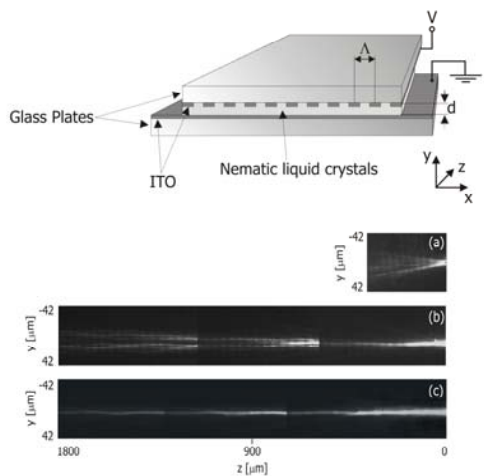

Figure 7.    (Top) Cell for liquid-crystal waveguide array: $\Lambda$, array period, $d$, cell thickness, $V$, applied voltage, ITO, indium tin oxide electrodes. (Bottom) Experimental results with light propagation at 1064 nm wavelength: (a) one-dimensional diffraction at $V = 0$ V, (b) low-power discrete diffraction at $V = 1.2$ V and 20 mW input power, (c) discrete nematicon at $V = 1.2$ V and 35 mW input power. The residual coupling and transverse-longitudinal power fluctuations shown are due partly to the limits of the camera and partly to nonuniformities in the sample (Fratalocchi, Assanto, Brzdakiewicz, Karpierz [2004]).

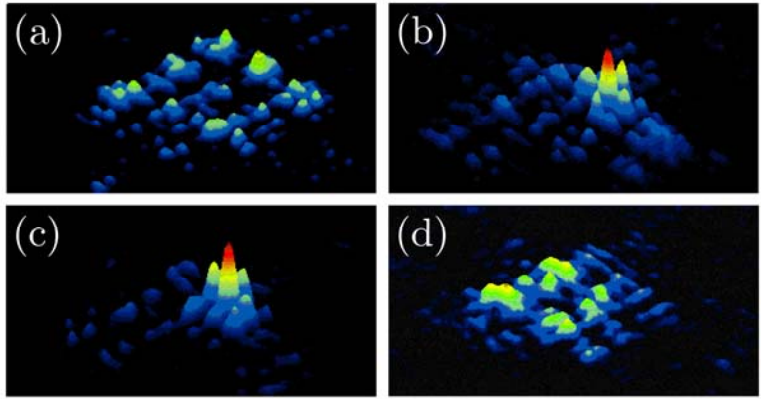

Figure 8.    Experimental results presenting the propagation of a probe beam launched into a single waveguide at normal incidence. (a) Intensity structure of the probe beam at the exist face of the crystal, displaying discrete diffraction at low nonlinearity (200 V). (b) Intensity structure of a probe beam at the exit face of the crystal, displaying an on-axis lattice soliton at high nonlinearity (800 V). (c) Interferogram, showing constructive interference of peak and lobes between the soliton and a plane wave. (d) Reduction of probe intensity by a factor of 8, at the same voltage as (b) and (c), results in recovery of discrete diffraction pattern (Fleischer, Segev, Efremidis, Christodoulides [2003]).

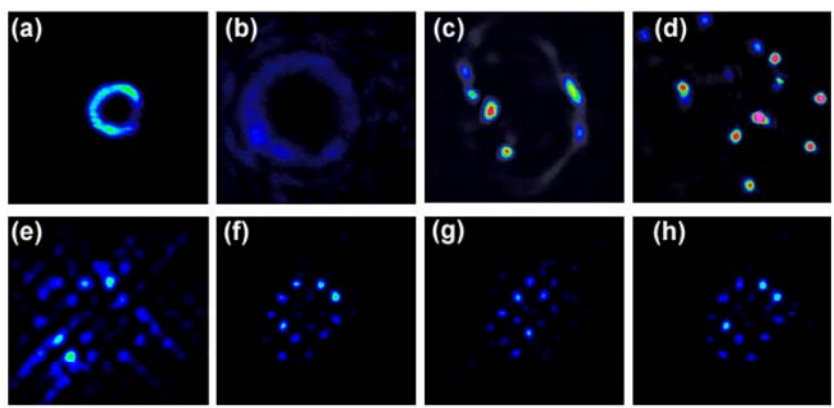

Figure 9.    Experimental results of charge-4 vortex propagating without (top) and with (bottom) the 2D lattice for different applied electric fields. (a) input, (b) linear diffraction, (c) output at a bias field of 90 V/mm, (d) output at 240 V/mm, (e) discrete diffraction at 90 V/mm, (f)-(h) discrete trapping at 240 V/mm. From (f) to (h), the vortex was launched into lattice from off-site, to on-site, and then back to off-site configurations. Distance of propagation through the crystal is 20 mm; average distance between adjacent spots in the necklace is 48 $\mu$m (Yang, Makasyuk, Kevrekidis, Martin, Malomed, Frantzeskakis, Chen [2005]).

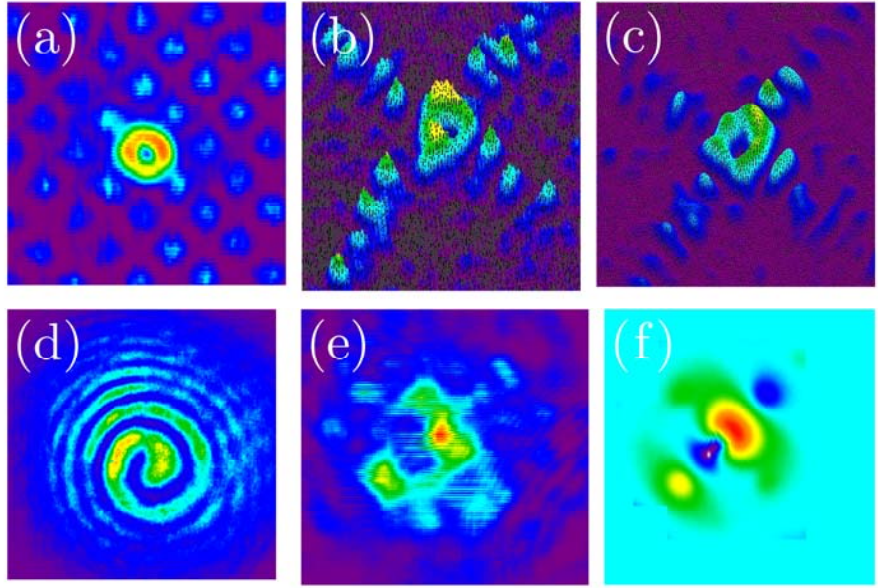

Figure 10.   Experimental observation of gap vortex lattice soliton. (a) Intensity distribution of the input vortex-ring beam photographed (for size comparison) on the background of the optically induced lattice. (b) Output intensity distribution of a low intensity ring beam experiencing linear diffraction in the lattice. (c) Output intensity distribution of a high intensity ring beam, which has evolved into gap vortex soliton in the same lattice as in (b). (d)-(f) Phase structure of gap vortex lattice soliton. The phase information is obtained by interference with a weakly diverging Gaussian beam. (d) Phase distribution of the input vortex-ring beam. (e) Output phase distribution of a high intensity ring-beam that has evolved into gap vortex soliton (interference pattern between the soliton whose intensity is presented in (c) and a weakly diverging Gaussian beam). (f) Numerical validation of the phase information in (e) (Bartal, Manela, Cohen, Fleischer, Segev [2005]).

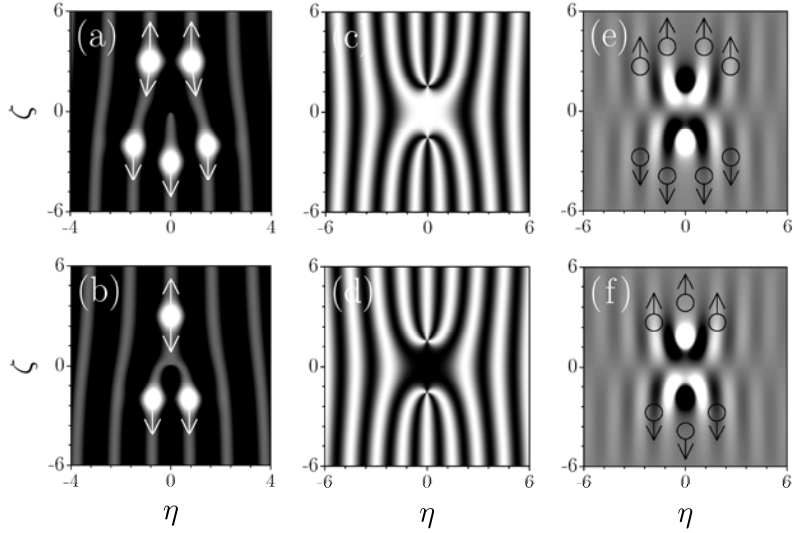

Figure 11.    (a) and (b) show lattices with single dislocations produced by the inter-
ference pattern of plane wave and wave with nested charge-1 vortex for
various vortex orientations. Arrows show direction of motion for solitons
placed into lattice channels in the vicinity of dislocation. Positive (c)
and negative (d) topological traps created with two vortices with
charges $m = \pm 3$. Map of $F_\zeta$ force for solitons launched in the vicinity
of positive (e) and negative (f) topological traps produced by two vor-
tices with charges $m = \pm 1$. Arrows show direction of soliton motion. In
(c)-(f) trap length $\delta\zeta = \pi$ (Kartashov, Vysloukh, Torner [2005b]).

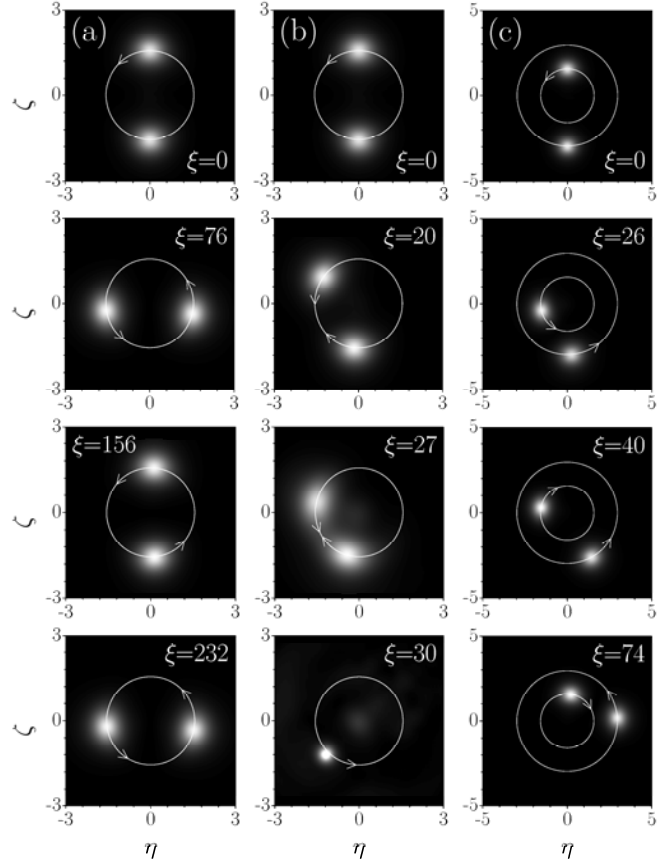

Figure 12.   (a) Interaction of out-of-phase solitons in the first ring of zero-order Bessel lattice. One of solitons is set into motion at the entrance of the medium by imposing the phase twist $\nu = 0.1$. (b) The same as in column (a) but for in-phase solitons. (c) Interaction of out-of-phase solitons located in first and second lattice rings. The soliton located in the first ring is set into motion by imposing the phase twist $\nu = 0.2$. All solitons correspond to the propagation constant $b = 5$ at $p = 5$. Soliton angular rotation directions are depicted by arrows (Kartashov, Vysloukh, Torner [2004c]).

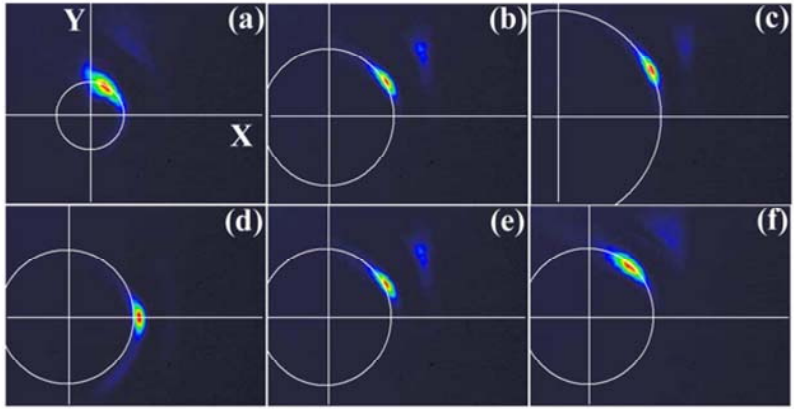

Figure 13. Soliton rotation in ring lattice. From (a) to (c), the probe soliton beam was aimed at the far right side of the first, second, and third lattice ring with the same tilting angle (about 0.4°) in the $y$ direction. From (d) to (f), the probe soliton beam was aimed at the same second lattice ring but with different tilting angles of 0°, 0.4°, and 0.6°, respectively (Wang, Chen, Kevrekidis [2006]).

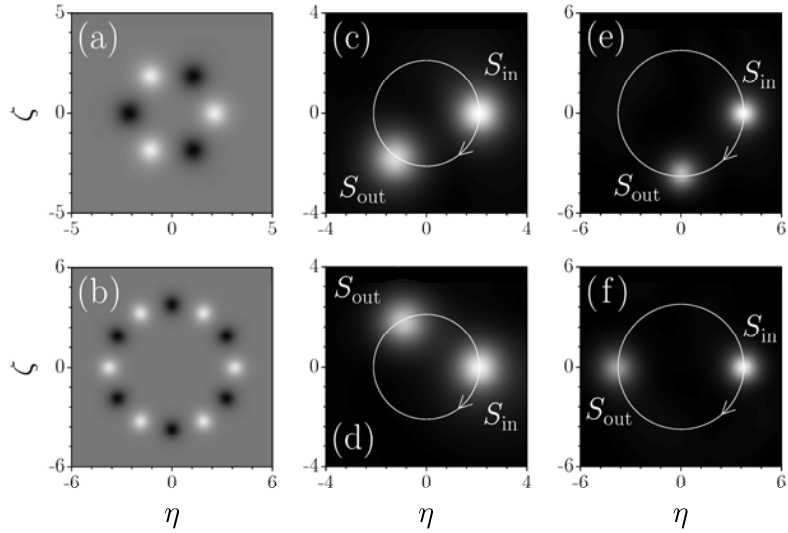

Figure 14.   Stable soliton complex supported by third-order Bessel lattice at $b = 3$
and $p = 20$ (a) and by sixth-order Bessel lattice at $b = 5$ and $p = 40$
(b). Azimuthal switching of single tilted soliton to second (c) and fourth
(d) channels of third-order Bessel lattice for tilt angles $\alpha_\zeta = 0.49$ and
$\alpha_\zeta = 0.626$ at $p = 2$ and input energy flow $U_{in} = 8.26$. Panels (e) and
(f) show switching to third and sixth channels of sixth-order lattice for
$\alpha_\zeta = 0.8$ and $\alpha_\zeta = 0.93$ at $p = 5$ and input energy flow $U_{in} = 8.61$.
Input and output intensity distributions are superimposed. Arrows show
direction of soliton motion and $S_{in}, S_{out}$ denote input and output soliton
positions (Kartashov, Egorov, Vysloukh, Torner [2004]).

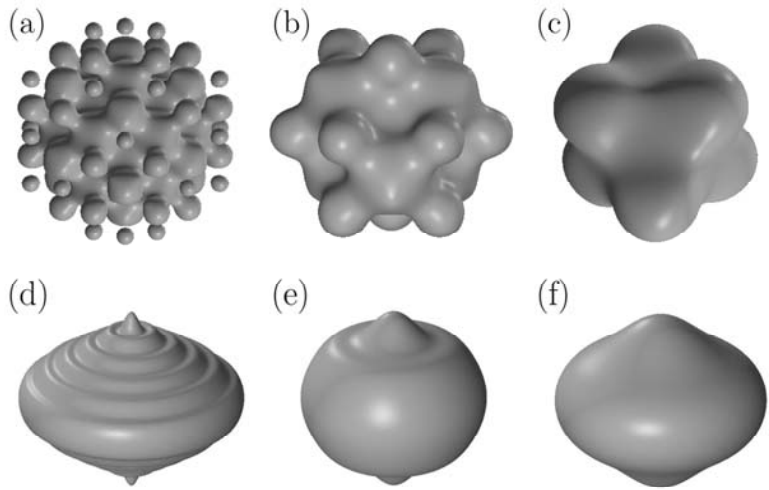

Figure 15.   Isosurface plots of (a) unstable and (b),(c) stable 3D solitons in 3D periodic lattice. In (a) soliton energy $U = 2.4$ and its peak amplitude $A = 1.8$, in (b) $U = 2.04$, $A = 2.2$, in (c) $U = 2.4$, $A = 3$. In (a)-(c) depth of periodic lattice $p = 3$. (d)-(f) Isosurface plots of stable 3D solitons supported by 2D Bessel lattice at (d)   $p = 5$,   $b = 0.11$,   (e) $p = 5.5$,   $b = 0.85$, and (f) $p = 6$,   $b = 2$ (Mihalache, Mazilu, Lederer, Malomed, Crasovan, Kartashov, Torner [2005]; Mihalache, Mazilu, Lederer, Malomed, Kartashov, Crasovan, Torner [2005]).

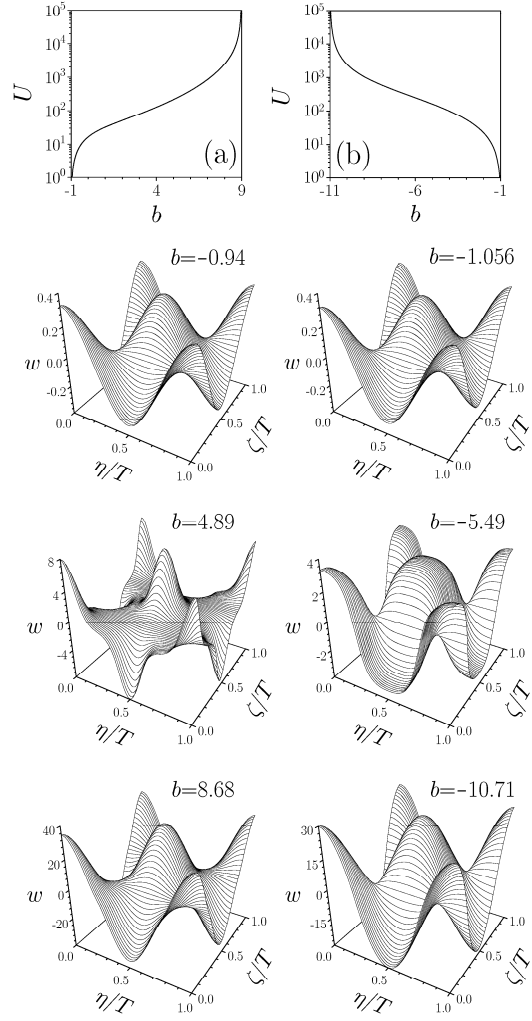

Figure 16. Cn-cn (a) and sn-sn (b) periodic waves in saturable medium with $S = 0.1$. Plots in the first row show energy flow concentrated within one wave period $T = 2\pi$ versus propagation constant. Surface plots in each column show evolution of wave shape with increase of energy flow $U$. Column (a) corresponds to focusing nonlinearity, column (b) corresponds to defocusing nonlinearity (Kartashov, Vysloukh, Torner [2003]).

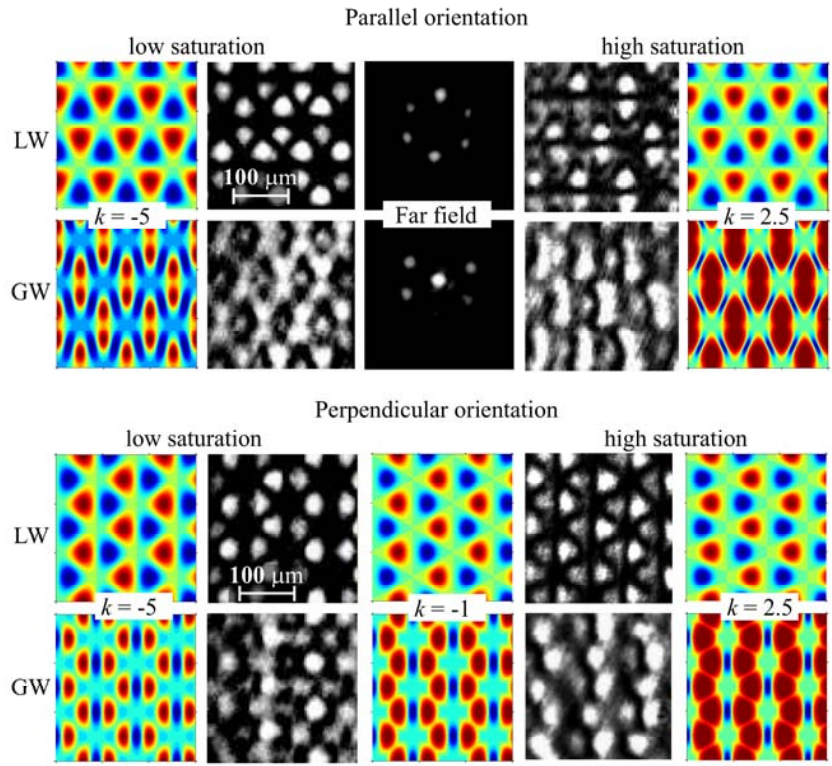

Figure 17. Theoretical (color) and experimental (grayscale) results for self-trapped triangular lattices created by interference of six plane waves, see far field images in the top panel. Two distinct orientations of the triangular lattices with respect to crystallographic $c$-axis are compared (top and bottom panels) as well as low (left) and high (right) saturation regimes. LW, lattice field (color) and intensity (grayscale) of the lattice wave. GW, calculated profiles of the refractive index (color) and measured intensity of the probe linear wave guided by the lattice (grayscale) (Desyatnikov, Sagemerten, Fisher, Terhalle, Trager, Neshev, Dreischuh, Denz, Krolikowski, Kivshar [2006]).

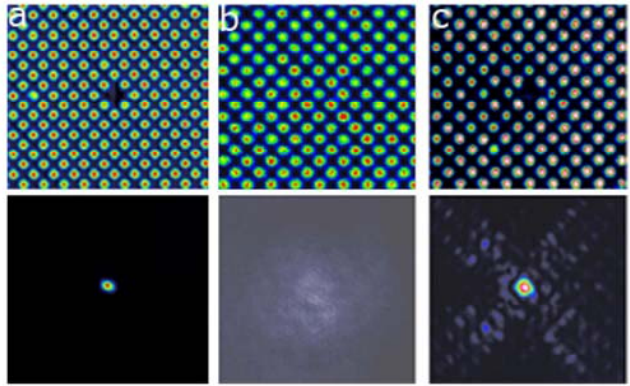

Figure 18.   Top panel: Intensity pattern of a 2D optically induced lattice with a single-site negative defect at crystal input (a) and output (b), (c). The defect disappears in the linear regime (b) but can survive with weak nonlinearity (c) after propagating through a 20-mm-long crystal. Bottom panel: Input (a) and output (b), (c) of a probe beam showing bandgap guidance by the defect (c) and normal diffraction without the lattice (b) under the same bias condition (Makasyuk, Chen, Yang [2006]).